\PassOptionsToPackage{expansion=false}{microtype}
\documentclass[sigconf,balance=false]{acmart}
\usepackage{popets}
\usepackage{color}
\usepackage{booktabs}
\usepackage{multirow}
\usepackage{array}
\usepackage{graphicx}
\usepackage{amsmath}
\usepackage{url}
\usepackage{xspace}
\usepackage[dvipsnames]{xcolor}
\usepackage{tikz}
\usepackage{soul}
\usepackage{float}
\usetikzlibrary{positioning, arrows.meta}
\usepackage{subcaption}
\usepackage{enumitem}
\usepackage{makecell}
\usepackage{soul}  
\usepackage{caption}
\usepackage{placeins}
\usepackage{microtype}   
\usepackage[normalem]{ulem}
\newcommand{\rev}[1]{#1}

\newcommand{\cyan}[1]{\textcolor{teal}{#1}}

\renewcommand{\cyan}[1]{#1}
 
\newcommand{\para}[1]{\vskip 4pt\noindent\textbf{#1}\hskip .05in}

\setcopyright{popets}
\copyrightyear{2027}
\acmYear{2027}
\acmVolume{2027}
\acmNumber{X}
\acmDOI{XXXXXXX.XXXXXXX}
\acmISBN{}
\acmConference{Proceedings on Privacy Enhancing Technologies}
\begin{document}

\title{Online Safety Regulation Increases Attention to VPNs: Privacy Implications of the UK Online Safety Act}

\author{Dhyey Mehta}
\affiliation{%
  \institution{University of Edinburgh}
  \city{Edinburgh}
  \country{UK}
}
\email{D.M.Mehta@sms.ed.ac.uk}

\author{Eldar Jalilzade}
\affiliation{%
  \institution{Newcastle University}
  \city{Newcastle upon Tyne}
  \country{UK}
}
\email{E.Jalilzade2@newcastle.ac.uk}

\author{Maksim Kalameyets}
\affiliation{%
  \institution{Newcastle University}
  \city{Newcastle upon Tyne}
  \country{UK}
}
\email{Maksim.Kalameyets@newcastle.ac.uk}

\author{Rebecca Owens}
\affiliation{%
  \institution{Durham University}
  \city{Durham}
  \country{UK}
}
\email{rebecca.owens@durham.ac.uk}

\author{Marc Juarez}
\affiliation{%
  \institution{University of Edinburgh}
  \city{Edinburgh}
  \country{UK}
}
\email{marc.juarez@ed.ac.uk}

\author{Stergios Aidinlis}
\affiliation{%
  \institution{Durham University}
  \city{Durham}
  \country{UK}
}
\email{stergios.aidinlis@durham.ac.uk}

\author{Lei Shi}
\affiliation{%
  \institution{Newcastle University}
  \city{Newcastle upon Tyne}
  \country{UK}
}
\email{lei.shi@newcastle.ac.uk}

\author{Tuğrulcan Elmas}
\affiliation{%
  \institution{University of Edinburgh}
  \city{Edinburgh}
  \country{UK}
}
\email{tugrulcan@ed.ac.uk}
\renewcommand{\shortauthors}{Mehta et al.}

\begin{abstract}
Governments worldwide are increasingly regulating digital platforms to reduce online harms, particularly those affecting children. However, access restrictions can alter user behaviour and introduce new privacy and security risks. The UK Online Safety Act (OSA), passed in October 2023, illustrates this trend: it extends age-assurance and safety requirements to social media, search, and pornography services, and rolled out in phases. Ofcom's illegal content enforcement duties came into force in March 2025, and mandatory age verification for adult content took effect in July 2025. This phased rollout enables real-time observation of behavioural responses to regulation. To address this, we analyse Reddit discourse across VPN and UK Politics communities and conduct a privacy-policy risk analysis of 69 unique VPN services.

\cyan{We find that the behavioural response is concentrated overwhelmingly {at the July 2025 deadline, when platforms hosting pornographic content were required to deploy age checks}. UK VPN search interest on Google increased by $+147\%$ at this deadline. Similarly, UK-resident users' activity in VPN-related subreddits increased by {$+145\%$}. {Their} regulatory- or privacy-related VPN posts and comments rose by 1,265\% at this deadline. UK Politics communities show the same concentration at a larger magnitude, with OSA-related political discourse rising by $+1481\%$ at the age-verification deadline. These effects were far smaller or statistically indistinguishable from pre-existing trends at Royal Assent and the illegal-harms enforcement deadline, indicating that the deployed age checks drove the response. Users primarily frame this response around privacy, surveillance, and distrust of age-verification intermediaries rather than simple access-seeking{, with near-zero genuine pro-OSA sentiment across two independent classifiers. Several users noted that those least able to pay for reputable VPNs are the ones most likely to turn to free services that monetise their data}. Search attention increases across all disclosed privacy-risk categories, with no evidence of a shift toward higher-risk VPN providers. Crucially, after a full year, this attention is still elevated, arguing against a temporary news-cycle reaction. Thus, online safety regulation may create secondary privacy costs without disproportionately directing attention toward higher-risk VPNs.}
\end{abstract}


\maketitle
\section{Introduction}
\label{sec:introduction}
Age assurance is increasingly central to digital regulation across many jurisdictions, aiming to prevent children from accessing age-restricted and harmful online products and services such as gambling and pornography~\cite{livingstone2024children, persson2024age}. Australia, for example, has legislated a social-media minimum age~\cite{AustraliaSocialMediaMinimumAgeAct2024}, while the EU is advancing similar requirements under Article~28(1) DSA. The United Kingdom provides one of the broadest examples of early implementations, through the Online Safety Act 2023~\cite{UKOnlineSafetyAct2023}, which policymakers claimed would make `the UK the safest place to be online in the world'~\cite{DSIT2023OnlineSafetyActLaw}. Instead of limiting regulation to adult-content websites, the Act adopts a differentiated duty-of-care approach for regulated user-to-user, search, and pornography services to mitigate the risks of illegal and harmful content and of children's exposure to it~\cite{law2024effective}. In parallel, the deployment of age-assurance mechanisms requiring identity documents to verify age has accelerated~\cite{woodley2025australian}.

Understanding how users respond to these measures is essential for evaluating their real-world privacy and security consequences.
Age verification may require sensitive information, including facial scans, identity documents, or payment details; individuals who distrust these systems may seek to obtain access without having to submit personal data to age-verification intermediaries.
A foreseeable response is increased use of privacy-enhancing technologies, such as Tor and VPNs: by routing traffic through a server in a different jurisdiction, these technologies can mask the user's location, allowing them to bypass the geolocation mechanisms that age verification systems rely on.

However, VPN use carries its own risks. First, children may use the same circumvention tools to bypass OSA protections, reducing the effectiveness of the regulatory scheme. Second, users who adopt VPNs for privacy reasons may expose themselves to new risks if they choose free or less reputable providers that log, track, retain, or share user data. For example, in 2020, a 1.2TB leak revealed that seven free VPN services were not honouring their no-log policies exposing personal and security-sensitive information, such as passwords and email addresses, of up to 20M users~\cite{pcmag2020vpn}.
Therefore, while the Act intends to reduce some online harms, it may inadvertently expose users to new privacy risks, especially if it nudges users toward less private alternatives.

As age-assurance and platform-access regimes expand globally, the UK OSA provides an early case study for studying user behaviour under child-safety rules enforced through access-control systems. We address three research questions, outlining the methodology and the key contributions the paper makes.

\begin{description}[leftmargin=*,topsep=2pt,parsep=2pt]

  \item[\textbf{RQ1.}] \textit{What is the causal impact of the OSA on the volume of VPN-related and politics discourse on Reddit and Google Trends?}
        
  We use Bayesian Structural Time Series (BSTS) to analyse Reddit post and comment counts, both raw and OSA-related, before and after each milestone, as well as Google Trends search volumes. Our analysis reveals OSA VPN discourse rises sharply at that July 2025 age-verification deadline, with no significant change at Royal Assent or the March 2025 enforcement deadline. The effect is largest among likely UK-resident authors. UK VPN-related Google Trends search interest follows the identical pattern, confirming the Reddit findings with an independent measure.

  \item[\textbf{RQ2.}] \textit{How is the OSA-related discussion framed? Are users pro or against the Act, and what arguments do they advance?}

  We use topic modelling and LLM-assisted summarisation on Reddit content to identify key narratives: why users prefer VPNs over facial age verification, their expressed concerns, and their views on the legislation. Our findings show that OSA discourse is mainly critical. Users present VPN use as privacy-preserving resistance to identity checks, facial age estimation, surveillance, and perceived overreach, rather than simply a means to access restricted content.

  \item[\textbf{RQ3.}] \textit{{Around the OSA milestones, did search attention shift toward VPN providers with higher policy-disclosed privacy risk?}}
  
   We (a) analyse 69 unique VPN services using archived privacy-policy pages, (b) extract markers of logging, tracking, retention, sharing, and policy ambiguity to classify each provider's disclosed privacy risk, and (c) compare Google Trends attention across low, medium, and high disclosed-risk providers. We find that VPN search attention increased across all three risk categories, but the relative share of attention remained broadly stable. The OSA increased VPN-related attention overall without shifting it disproportionately toward higher disclosed-risk providers.

\end{description}

\section{Motivation and Background}
\label{sec:background}

\para{The Online Safety Act (OSA).}
The UK's OSA was introduced following the recommendations of the DCMS Online Harms 2019 White Paper\cite{DCMSHomeOfficeOnlineHarmsWhitePaper2019}. The Act received Royal Assent on 26 October 2023 and established a regulatory framework under which providers of regulated user-to-user and search services are subject to duties of care and are supervised and enforced by Ofcom. Those duties include obligations to assess and mitigate risks of illegal content and, for services likely to be accessed by children, to identify and reduce risks of harm to children. Crucially, a service is in scope of the Act if it targets UK users, has a significant UK user base, or is accessible in the UK and presents a material risk of significant harm to individuals there. In practice, this means that the largest social media, search and pornography platforms are covered under the Act \cite{mcglynn2024pornography}.

This study focuses on three key milestones to assess the Act's impact on VPN attention and privacy discussions. 
\begin{enumerate}[leftmargin=*,topsep=1pt]
    \item {\bf Enacted.} In October 2023, \textit{Royal Assent}  formally enacted the regime and signalled that online access controls would become central to UK platform regulation.
    \item {\bf In Force.} From March 2025, Ofcom's illegal harms regime became enforceable, requiring regulated services to conduct risk assessments and address illegal content.
    \item {\bf In Effect.} From 25 July 2025, services hosting or allowing pornographic content, as well as those subject to children's safety duties, were required to implement robust age assurance measures, such as age verification or estimation, to prevent children from accessing inappropriate content~\cite{ofcom2025agechecksinforce}. Independently reported compliance dates cluster tightly around this deadline: Discord from 21 July~\cite{discord2025osa}, Aylo (Pornhub/YouPorn/RedTube) from 24 July~\cite{aylo2025ageassurance}, Reddit from 25 July~\cite{eff2025redditrollout}, and Bluesky by 25 July~\cite{forbes2025bluesky}.
\end{enumerate}
\noindent

\para{VPN Adoption and Internet Censorship.} Internet censorship is still prevalent today~\cite{ccetinkaya2025state} causing people to adopt VPNs, but prior work suggests that VPN adoption is best understood as an event-driven
response to new access frictions. Hobbs and Roberts describe a ``gateway effect'': when states
suddenly block high-demand services, users resort to circumvention tools to
preserve existing habits, incidentally gaining access to other already-blocked
platforms -- in China, the 2014 Instagram block pushed millions to acquire VPNs
and reach long-blocked services such as Twitter and Facebook~\cite{hobbs2018sudden}.
Similar dynamics appear elsewhere: in Iran, VPN use reached an estimated 64\% of
internet users in 2023 after the 2022 protests, alongside a 2024 ban on
unlicensed VPNs~\cite{freedomhouseIran2024,internetSocietyIran2024,iranwire2025vpn}.
Adoption is uneven, however, deterred by cost, trust, service quality, and legal
risk~\cite{freedomhouseRussia2024,levadaRussia2025,xue2024bridging}.

These cases identify the key mechanism: when regulation creates visible access
frictions around widely used services, VPN use can become normalised quickly.
{At the same time, adoption is socially and demographically uneven. Decisions
to adopt or abandon VPNs reflect both emotional and practical considerations,
including privacy concerns, trust, cost, and usability~\cite{namara2020emotional}.}
Younger and more digitally literate users may be more likely to adopt
circumvention tools, while others may be deterred by cost, trust concerns,
degraded service quality, or uncertainty about legal and platform
risks~\cite{xue2024bridging,levadaRussia2025,freedomhouseRussia2024}. Provider
choice adds a further complication, as users must assess competing provider
claims about privacy, security, and performance~\cite{ramesh2023best}. The OSA
offers a distinct case because the access friction is justified by child-safety
regulation rather than conventional political censorship, but the behavioural
pathway may be similar (turning to VPNs to preserve access and privacy).

\para{Reddit as a Data Source for Policy Research.}
Reddit has been widely used to study public discourse on policy and social issues at
scale~\cite{proferes2021studyingreddit}, with data collected via
Pushshift~\cite{baumgartner2020pushshift} and its successor Arctic Shift. Its subreddit
structure lets us observe the same legislation discussed in two distinct community types:
VPN communities and UK Politics communities.
Prior work establishes the methodological foundation for our approach. Kim et
al.~\cite{kim2025ubi} and Xu et al.~\cite{xu2024chatgpt} show that combining LDA topic
modelling with LLM-assisted labelling given parent post context produces more reliable
topic assignments than LDA alone, motivating
our hybrid pipeline for RQ2. Saha et al.~\cite{saha2024observer} apply CausalImpact to longitudinal
Facebook data, finding that a single discrete event changed posting behaviour
and linguistic attributes, demonstrating
that BSTS isolates event-driven shifts from underlying trends in noisy social media
timeseries. This is the core methodological challenge for RQ1: isolating the OSA's causal effect from
the organic growth in VPN and privacy discourse that predates the Act.

\flushbottom

\section{Methodology Overview}
\label{sec:method}

\begin{figure}[t]
    \centering
    \includegraphics[width=\linewidth]{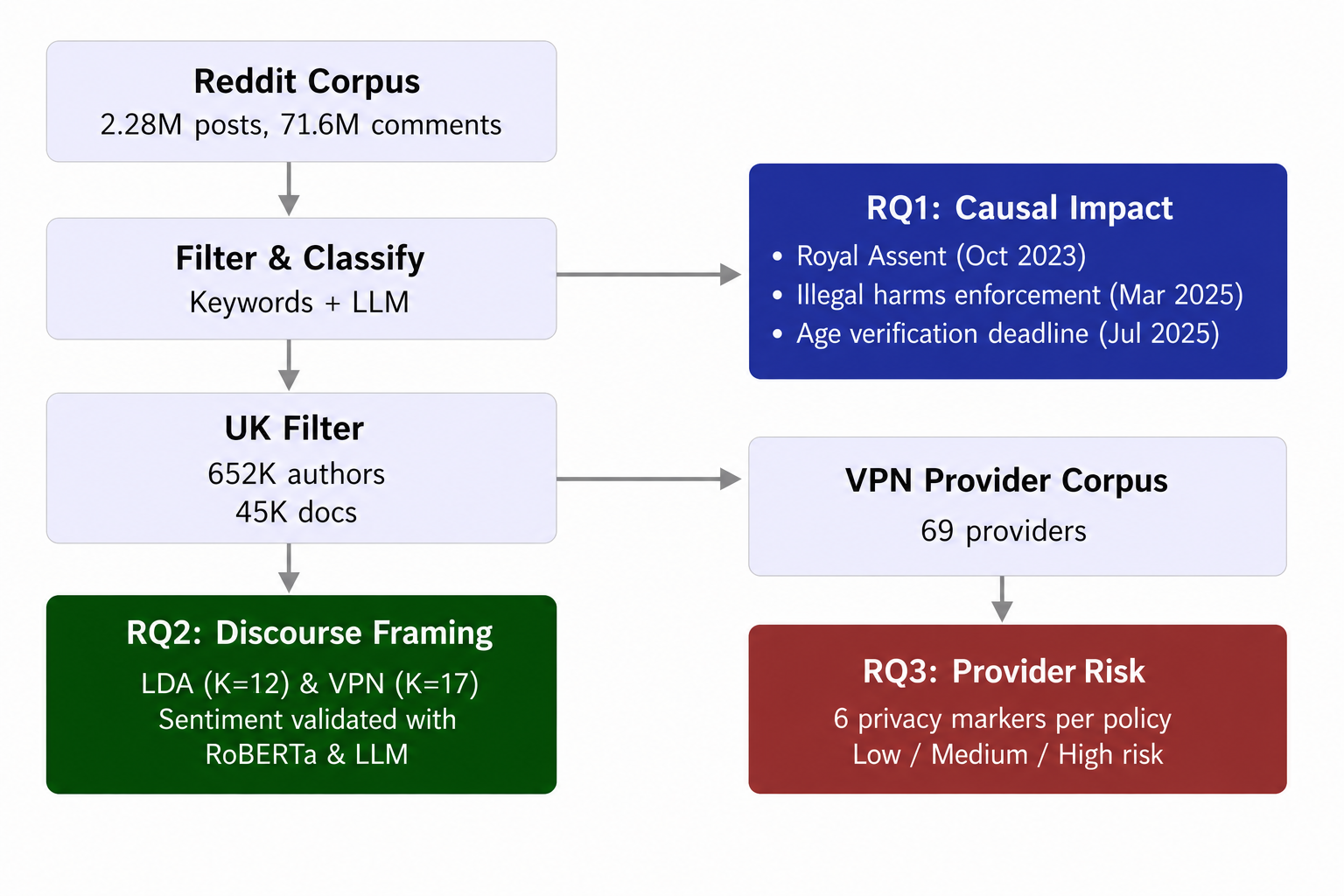}
    \caption{Overview of the methodology pipeline.}
    \label{fig:pipeline}
\end{figure}

    Our methodology is designed to trace the full pathway from regulation to user response and then to downstream privacy risk. Figure \ref{fig:pipeline} summarises this workflow. We begin by constructing a Reddit corpus covering VPN-related and UK Politics communities from October 2021 to October 2025. Because the raw corpus contains many posts and comments that mention VPNs, privacy, or regulation only indirectly, we first apply broad keyword filters and then use an LLM relevance classifier to identify documents that are actually about the OSA, age assurance, VPN use, circumvention, censorship, or related privacy concerns. This produces a classified corpus of 78,689 OSA-relevant documents, which forms the basis for the RQ1 and RQ2 analyses. We additionally identify potential UK-resident authors from their UK-related subreddit activity to better represent the target population.

    For RQ1, we use the Reddit corpus to measure whether OSA milestones changed the volume of online discussion and information-seeking behaviour. We construct weekly time series for posts, comments, and users, both across all authors and within the UK-resident subset. We then apply Bayesian Structural Time Series models to estimate what each series would have looked like without the relevant OSA milestone. Comparing the observed series with this counterfactual allows us to estimate the effect of Royal Assent, the illegal-harms enforcement milestone, and the age-assurance deadline. We also analyse UK Google Trends search interest for VPN-related terms as an independent signal outside Reddit. This step shows whether the Act produced measurable behavioural displacement toward VPN-related attention.

    For RQ2, we move from volume to meaning. We restrict the analysis to the 45,438 OSA-relevant documents written by likely UK-resident authors and use topic modelling, LLM-assisted topic interpretation, and sentiment analysis to examine how users discuss the Act. This part of the pipeline identifies the main narratives in VPN and UK Politics communities, including concerns about facial age verification, identity checks, surveillance, censorship, and distrust of age-assurance intermediaries.

    For RQ3, we examine the privacy implications of this increased VPN attention. We analyse archived privacy-policy pages for 69 unique VPN services, classify each service into a privacy-risk category, and compare monthly UK Google Trends attention across risk categories. We then link these categories to saved monthly UK Google Trends attention for VPN provider names. This final step asks whether increased VPN attention was concentrated among higher disclosed-risk providers or instead reflected a broader rise in attention across the VPN market.

\subsection{Reddit Corpus and Preprocessing}
\label{subsec:shared_corpus}
\label{sec:data}
\label{sec:classification}

    We construct a Reddit corpus used across the causal and discourse analyses using Project Arctic Shift~\cite{arcticshift2022}. The corpus spans October 2021 to October 2025, covering a four-year window that includes a 24-month pre-intervention period before Royal Assent on 26 October 2023 and the subsequent OSA enforcement timeline through the age-verification deadline. We excluded data from \texttt{AutoModerator} (a bot account) and \texttt{[deleted]} authors.
    VPN subreddits were selected on the basis of relevance to VPN use, privacy, and circumvention, and sufficient posting volume for time-series analysis. UK Politics subreddits were selected to cover mainstream UK political discussion across party-affiliated, regional, and general public-affairs communities, given our interest in OSA-related political discourse.
    We focus on two subreddit groups: VPN-related communities and UK Politics communities. We additionally collect data from a set of UK geographic subreddits used for author geo-filtering. Table~\ref{tab:subreddits} lists the selected subreddits.

\begin{table}[t]
\centering
\caption{Subreddits selected for analysis and their statistics.}
\label{tab:subreddits}
\footnotesize
\setlength{\tabcolsep}{2pt}

\begin{tabular}{p{0.15\columnwidth}p{0.53\columnwidth}p{0.12\columnwidth}p{0.13\columnwidth}}
\toprule
\textbf{Group} & \textbf{Subreddits} & \textbf{Posts} & \textbf{Comments} \\
\midrule
VPN (11) &
r/VPN, r/privacy, r/PrivacyGuides, r/degoogle, r/ProtonVPN, r/nordvpn, r/MullvadVPN, r/Surfshark, r/PrivateInternetAccess, r/Windscribe, r/censorship
& 243,150 & 2,275,677 \\

UK Politics (18) &
r/ukpolitics, r/UnitedKingdom, r/AskUK, r/CasualUK, r/BritishProblems, r/LegalAdviceUK, r/Scotland, r/Wales, r/NorthernIreland, r/England, r/labour, r/LabourUK, r/LibDem, r/Tories, r/GreenAndPleasant, r/FreeSpeech, r/uklaw, r/UKGreens
& 2,040,904 & 69,361,713 \\

\midrule
\multicolumn{2}{l}{\textbf{Total}} & \textbf{2,284,054} & \textbf{71,637,390} \\
\midrule
\multicolumn{4}{p{0.95\columnwidth}}{\textbf{UK Geo-filter ({37}):} r/london, r/Edinburgh, r/glasgow, r/manchester, r/Birmingham, r/bristol, r/Cardiff, r/Leeds, r/Liverpool, r/Belfast, r/AskLondon, r/LondonUnderground, r/londonersr4r, r/NorthEastUK, r/UKJobs, r/uklandlords, r/drivingUK, r/policeuk, r/CoronavirusUK, r/heyUK, r/BirminghamUK, r/MakeFriendsUK, r/MakeMoneyInUK, r/UK\_Food, r/UK\_Pets, r/UK\_News24, r/UKcoins, r/SkilledWorkerVisaUK, r/HumanResourcesUK, r/ContractorUK, r/ActuaryUK, r/ArchitectsUK, r/AmazonFlexUK, r/AmexUK, r/apprenticeuk, r/JustEatUK, r/McDonaldsUK} \\
\bottomrule
\end{tabular}
\end{table}

    \noindent
    \para{Keyword pre-filtering and LLM classification.} The raw Reddit corpus contains many posts and comments that mention VPNs, privacy, regulation, or UK politics only indirectly. To retain documents relevant to VPN use, internet regulation, and the Online Safety Act (OSA), we apply a two-step preprocessing pipeline: keyword pre-filtering followed by LLM relevance classification.

    First, we apply a broad keyword pre-filter to remove documents clearly outside the study scope and reduce the number of items passed to the LLM. For VPN subreddits, we use 44 VPN-related terms. For UK Politics subreddits, we use 27 OSA-related terms. The full keyword lists are provided in Appendix~\ref{appendix:keywords}. To capture cross-community discussion, we also apply the OSA keyword list to VPN subreddits and the VPN keyword list to UK Politics subreddits. These supplementary runs capture OSA discussion in VPN communities and VPN/circumvention discussion in political communities. Matches are deduplicated before classification.

    Second, we classify all keyword-matched documents using Gemini~2.5~Flash. This step is needed because keyword matching is intentionally broad and includes irrelevant material, such as VPN setup questions, streaming discussions, provider comparisons, or general political comments. We use separate prompts for VPN and UK Politics subreddits because the same terms can have different meanings across communities. The full prompts used for both subreddit groups are provided in Appendix ~\ref{appendix:prompts}.

  We validate the relevance classifier using a stratified random sample of 200 manually annotated documents, balanced across the VPN and UK Politics groups. For UK Politics, the classifier achieves a precision of 0.86, recall of 0.94, and F1 score of 0.90. Performance is weaker for VPN documents, with a precision of 0.58, recall of 1.00, and F1 score of 0.73. Given the low precision for the VPN group, we apply a second-stage classifier to filter out noisy documents. This increases precision to 0.77, while reducing recall to 0.70, with an F1 score of 0.73. Across both groups, the final classification pipeline achieves a precision of 0.83, recall of 0.84, and F1 score of 0.83. We use this classifier throughout the paper.

    Table~\ref{tab:corpus_composition} reports the number of keyword-matched and classified documents as a share of the full corpus, including the UK-resident-author breakdown. The table shows the scale of the keyword-filtered corpus and the amount of noise removed before constructing the RQ1 and RQ2 analysis datasets.

    \label{subsubsec:clf_results}

\begin{table}[h]
\centering
\small
\caption{Corpus composition: share of all raw VPN- and Politics-subreddit documents (Oct~2021--Oct~2025) falling into each filter category. ``UK'' = author flagged as UK-resident. ``Classified'' uses the relevance classifier.
}
\label{tab:corpus_composition}
\resizebox{\columnwidth}{!}{%
\begin{tabular}{lrr}
\toprule
\textbf{Category} & \textbf{VPN} & \textbf{Politics} \\
\midrule
Total documents         & 1,796,763 (100.00\%) & 53,001,179 (100.00\%) \\
Keyword-matched         & 317,575 (17.67\%)    & 542,539 (1.02\%) \\
Classified               & 20,954 (1.17\%)      & 57,735 (0.11\%) \\
From UK authors          & 153,544 (8.55\%)     & 42,095,371 (79.42\%) \\
UK \& keyword-matched     & 24,841 (1.38\%)      & 431,269 (0.81\%) \\
UK, matched \& classified & 2,477 (0.14\%)       & 43,291 (0.08\%) \\
\bottomrule
\end{tabular}}
\end{table}

    \para{UK-resident author filter.} For user-level and discourse-framing analyses, we identify potential UK-resident authors using a subreddit-based geographic filter. An author is treated as likely UK-resident if they posted in both a study subreddit group, either VPN or UK Politics, and at least one UK geographic subreddit listed in Table~\ref{tab:subreddits}. This yields 652,200 likely UK-resident authors, of whom 19,189 posted OSA-classified content.

\section{RQ1: Causal Impact on Online Discourse}
\label{subsec:rq1_method}
\label{subsec:bsts}

    RQ1 asks whether OSA milestones caused measurable changes in VPN-related and politics discourse on Reddit and in VPN-related search behaviour. We use Bayesian Structural Time Series (BSTS), implemented through the CausalImpact framework~\cite{brodersen2015inferring}, to estimate the causal effect of each milestone. CausalImpact fits a structural time series model to the pre-intervention period and predicts a counterfactual trajectory for the post-intervention period. The difference between the observed and counterfactual series is the estimated causal effect, reported as a relative effect with a 95\% credible interval and Bayesian posterior tail-area probability~$p$. We primarily use Reddit data which described in \ref{sec:data}.

    \para{Google Trends data.} In addition to Reddit activity, we also analyse UK VPN Google Trends search interest as an independent outcome to corroborate our findings. We collect weekly Google Trends data for the UK and US via the \texttt{pytrends} library over the same four-year window as the Reddit corpus, using \textit{vpn}, \textit{nordvpn}, \textit{expressvpn}, \textit{surfshark}, and \textit{mullvad} as query terms. These five terms were selected as the highest-search-volume VPN-related queries in the UK over the study window.

    \para{Intervention dates.} We employ the milestones from Section~\ref{sec:background}: Royal Assent (October 2023), the illegal harms enforcement deadline (March 2025), and the age-verification deadline (July 2025).

    \para{Primary specification.}
    \label{subsec:primary_spec}
    Every CausalImpact estimate uses no covariate (except Google Trends estimates), a growing pre-period (all data before the milestone), and an 8-week post-window beginning with the week containing the milestone. 4-week and 12-week post-windows are reported as robustness (\S\ref{sec:rq1_d_robustness}). UK-resident-author series use a 652,200-author footprint. Alternative author-residency thresholds are reported for robustness in Table~\ref{tab:app_k_robustness} (Appendix~\ref{appendix:uk_author_robustness}).

    \para{Outcome Variables and Covariates.} We analyse three outcomes, each with its own counterfactual strategy:
    \begin{itemize}[leftmargin=*]
\item \noindent\textbf{Reddit content (VPN and UK Politics).} Weekly post and comment
volume at three filtering levels (raw, keyword-matched, classified relevant),
for all authors and for likely UK-resident authors, benchmarked against
same-platform placebo series that strip the OSA mechanism (keyword-unmatched
content, non-UK authors, or both) to bound generic subreddit growth.
\item \noindent\textbf{User counts.} Distinct authors per week (unique, new, cumulative)
at the same filtering levels, separating broader participation from the same
accounts posting more; the placebo logic carries over.
\item \noindent\textbf{Google Trends UK search interest.} Weekly UK search volume for
VPN terms -- an outcome with neither Reddit's self-selection nor any classifier,
so agreement with the Reddit series is genuine corroboration. We use covariates
directly: US search interest (global VPN demand) and a 52-week lag (seasonality).
\end{itemize}

\subsection{Results}
\label{sec:rq1}

 We observe a step-change in Google trends and Reddit discourse during the Age Verification milestone, attributed to broader participation from likely UK resident users. Table~\ref{tab:content_effects} reports the effect on new users and content (posts+comments) for every series/milestone combination underlying the claims (posts/comments breakdown in Table~\ref{tab:content_effects_full}, Appendix~\ref{appendix:extra_robustness}, Google Trends results in Table~\ref{tab:trends}). We now describe findings in detail.

\begin{table}
\centering
\caption{CausalImpact relative effect on content volume (posts+comments combined) and on new-user count. Raw = all documents in the subreddit group; Classified = documents marked as OSA-relevant; (UK) restricts to likely UK-resident authors. $^{**}p<0.05$, $^{*}p<0.10$; ${\approx}$: not significant.}
\label{tab:content_effects}
{\setlength{\tabcolsep}{1.5pt}
\scriptsize
\begin{tabular}{>{\raggedright\arraybackslash}p{0.07\columnwidth}>{\raggedright\arraybackslash}p{0.095\columnwidth}>{\centering\arraybackslash}p{0.115\columnwidth}>{\centering\arraybackslash}p{0.115\columnwidth}>{\centering\arraybackslash}p{0.115\columnwidth}>{\centering\arraybackslash}p{0.115\columnwidth}>{\centering\arraybackslash}p{0.115\columnwidth}>{\centering\arraybackslash}p{0.115\columnwidth}}
\toprule
& & \multicolumn{2}{c}{\textbf{Oct 2023}} & \multicolumn{2}{c}{\textbf{Mar 2025}} & \multicolumn{2}{c}{\textbf{Jul 2025}} \\
\cmidrule(l{1pt}r{1pt}){3-4} \cmidrule(l{1pt}r{1pt}){5-6} \cmidrule(l{1pt}r{1pt}){7-8}
\textbf{Grp.} & \textbf{Filter} & Content & Users & Content & Users & Content & Users \\
\midrule
\multirow{4}{*}{VPN}
 & Raw & $-21\%^{**}$ & ${\approx}{-10\%}$ & $-21\%^{**}$ & $-30\%^{**}$ & $+75\%^{**}$ & $+63\%^{**}$ \\
 & Raw (UK) & ${\approx}{-6\%}$ & ${\approx}{-14\%}$ & $-15\%^{**}$ & ${\approx}{-6\%}$ & $+145\%^{**}$ & $+230\%^{**}$ \\
 & Classified & $-20\%^{**}$ & $-22\%^{**}$ & ${\approx}{-12\%}$ & $-15\%^{*}$ & $+328\%^{**}$ & $+303\%^{**}$ \\
 & Cls. (UK) & ${\approx}{-16\%}$ & ${\approx}{-16\%}$ & ${\approx}{-16\%}$ & ${\approx}{-23\%}$ & $+1265\%^{**}$ & $+1230\%^{**}$ \\
\midrule
\multirow{4}{*}{Politics}
 & Raw & ${\approx}{-4\%}$ & $-43\%^{**}$ & {${\approx}{-2\%}$} & ${\approx}{-2\%}$ & ${\approx}{-7\%}$ & ${\approx}+19\%$ \\
 & Raw (UK) & {${\approx}{-8\%}$} & $-48\%^{*}$ & {${\approx}{-2\%}$} & ${\approx}{-76\%}$ & {${\approx}{-4\%}$} & ${\approx}+124\%$ \\
 & Classified & ${\approx}+25\%$ & ${\approx}+10\%$ & $+65\%^{**}$ & $+27\%^{*}$ & $+1474\%^{**}$ & $+922\%^{**}$ \\
 & Cls. (UK) & ${\approx}+27\%$ & ${\approx}+13\%$ & $+60\%^{**}$ & ${\approx}+25\%$ & $+1481\%^{**}$ & $+863\%^{**}$ \\
\bottomrule
\end{tabular}}
\end{table}

\para{OSA milestones produce a step-change.} VPN-related activity does not increase consistently across the three OSA milestones. Raw posts and comments remains stable or decline following Royal Assent and Enforcement. 
Content classified as relevant to VPNs follows the same pattern. Both raw and classified activity show that VPN discourse moves decisively only at Age Verification. The effect is more pronounced when limited to classified content ($+328\%^{**}$) than for all raw content in the subreddit group ($+75\%^{**}$) (Table~\ref{tab:content_effects}; posts/comments breakdown in Table~\ref{tab:content_effects_full}).

\begin{figure*}[t]
\centering
\includegraphics[width=\textwidth]{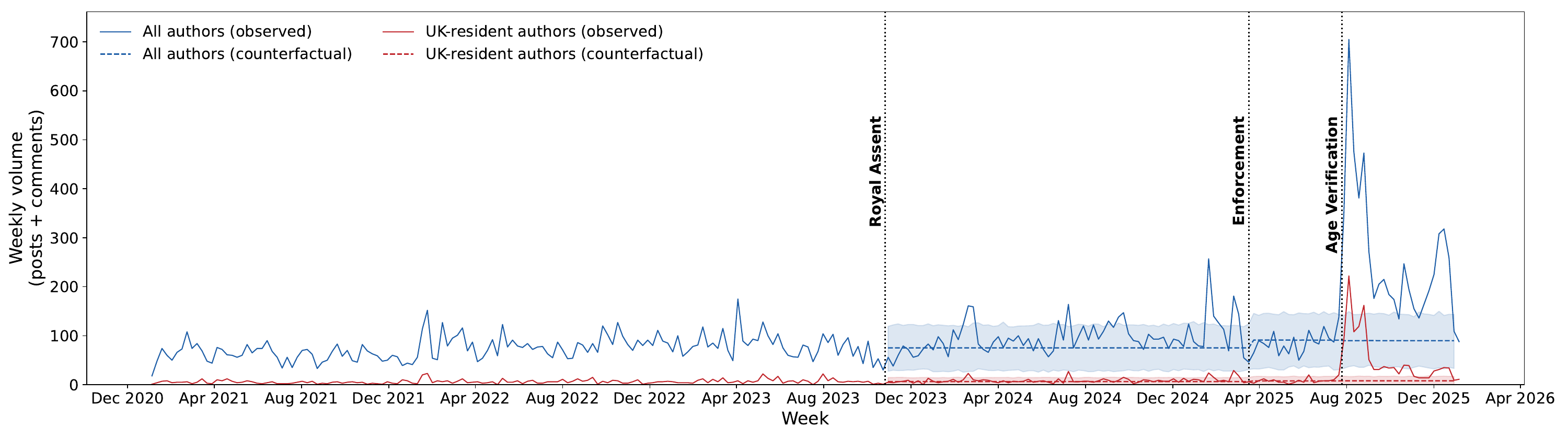}
\caption{CausalImpact
timeseries for VPN subreddits. Blue: all authors; red: UK-resident authors.
Solid lines show observed volume, dashed lines the BSTS counterfactual,
shaded regions the 95\% credible interval; vertical lines mark OSA
milestones. The data is extended to December 2025 for illustration purposes.}
\label{fig:rq1_vpn_classified}
\end{figure*}

As suggested by both the estimated effect size and the raw weekly counts in Figure~\ref{fig:rq1_vpn_classified}, activity in VPN-related subreddits declined around the time of Royal Assent in October 2023.
Inspecting the data, we find that this is due to a spurious effect on r/privacy caused by moderator decisions; moderators posted publicly on 13 October 2023 that the subreddit was switching to links-only submissions for the rest of the month because of a moderator-capacity shortage, and confirmed the reversal on 2 November, the exact week volume recovers (Appendix~\ref{appendix:royal_assent_null}). The same subreddit also observed a surge of activity prior to Enforcement due to the high volume of discussions on age-verification laws globally (US state laws, EU). Neither pattern coincides with anything to do with UK's OSA; this is consistent with, and gives a concrete mechanism for, treating Royal Assent and Enforcement as failing identification (Appendix~\ref{appendix:royal_assent_null}). Meanwhile, r/privacy contributed the most to the effect post Age Verification with a $48.2\%$ increase in OSA related posts and comments. This is more than all VPN-provider subreddits combined ($32.9\%$); Table~\ref{tab:av_subreddit_vpn}, and inspecting the top posts and comments verified these are related to the UK's OSA.

Classified posts and comments on UK Politics subreddits increase following all three milestones. The estimates for Royal Assent and Enforcement, however, should not be interpreted causally. This is first because the absolute increase in posts and comments is marginal relative to Age Verification ($+224$ extra posts/comments at Royal Assent and $+718$ at Enforcement, versus $+20{,}310$ at Age Verification). Additionally, placebo tests show that estimates increase when milestones are replaced with earlier dates. The estimated increases at Royal Assent and Enforcement therefore appear to capture pre-existing trends rather than effects of the milestones themselves (Table~\ref{tab:app_placebo_ra}), and inspecting the classified content directly confirms this for both: the Royal Assent rise is sustained news coverage of the Online Safety Bill's own final Parliamentary passage on 19 September 2023, more than a month before Royal Assent, and the Enforcement rise traces to Ofcom's publication of its Illegal Harms Codes of Practice on 16 December 2024, almost exactly three months before Enforcement (Appendix~\ref{appendix:royal_assent_null}).

Meanwhile, Age Verification is not subject to the same concern. Classified OSA-related discourse increases by $+1474\%^{**}$ (content; Politics, Table~\ref{tab:content_effects}). For VPN, placebo tests using pseudo-treatment dates ten and twelve weeks before Age Verification find no effect. The estimate at eight weeks before is weak and substantially smaller ($+41\%$, $p=0.083$). The result is also robust to alternative treatment dates: moving the treatment date across the four dates on which different platforms introduced age checks changes the estimated effect only from $+1248\%$ to $+1261\%$ (Table~\ref{tab:app_placebo_ra}c) for VPN-UK content volume. Thus, the same pre-trend placebo test that undermines a causal interpretation of the earlier milestones supports one for Age Verification for VPN (Appendices~\ref{appendix:royal_assent_null} and~\ref{appendix:age_verification_manipulation_check}). For Politics, the placebo test itself shows a significant pre-trend eight weeks before Age Verification (Table~\ref{tab:app_placebo_ra}c), so this specific check does not clear it as cleanly; there, the case rests instead on the milestone effect being an order of magnitude larger than the pre-trend, its concentration in OSA-relevant content rather than the subreddit group generally, and reading the classified content directly confirming genuine engagement rather than a classifier artefact (Appendix~\ref{appendix:age_verification_manipulation_check}). The timing and magnitude of the estimates therefore provide strong evidence that the implementation of age-verification requirements caused a substantial increase in attention to VPNs on Reddit. For Politics, the same magnitude and content-level evidence support a substantial rise in OSA-related political discourse at Age Verification, though the pre-trend placebo test's failure there means this falls short of the same clean causal identification VPN has.

\para{VPN attention concentrates among UK authors.} UK-authored content rises much more steeply than all-author content ($+1265\%^{**}$ vs. $+328\%^{**}$, a $3.9\times$ difference; Figure~\ref{fig:rq1_vpn_classified}). Politics behaves differently: the UK and all-author effects at Age Verification are almost identical ($+1481\%^{**}$ vs. $+1474\%^{**}$, Table \ref{tab:content_effects}, Figure ~\ref{fig:rq1_politics_classified_companion}). This is likely because 80\% of total posts and comments in Politics subreddits are already sourced from likely UK authors as shown in Table \ref{tab:corpus_composition}.

\para{New users participate on Reddit.} The effect can be attributed to more people joining the discussion or the same people speaking louder by repeated posting. Table~\ref{tab:content_effects} suggests the former: new (first-time) VPN authors are flat or declining at Royal Assent and Enforcement ($-22\%^{**}$, $-15\%^{*}$ all authors; ${\approx}{-16\%}$, ${\approx}{-23\%}$ UK), then rise sharply at Age Verification ($+303\%^{**}$ all authors, $+1230\%^{**}$ UK). Politics new users show a comparable rise at Age Verification ($+922\%^{**}$ all authors, $+863\%^{**}$ UK), matching the shared UK/non-UK pattern from the previous claim. Placebo series confirm this is genuine engagement rather than ordinary subreddit growth, and that the effect is UK-concentrated for VPN but shared across UK and non-UK authors for Politics (full derivation in Appendix~\ref{appendix:uk_author_robustness}). Full unique/new/cumulative user-count results, including the placebo rows behind this claim, are in Table~\ref{tab:users} (Appendix~\ref{appendix:age_verification_manipulation_check}).

\begin{figure}[tb]
\centering
\includegraphics[width=\columnwidth]{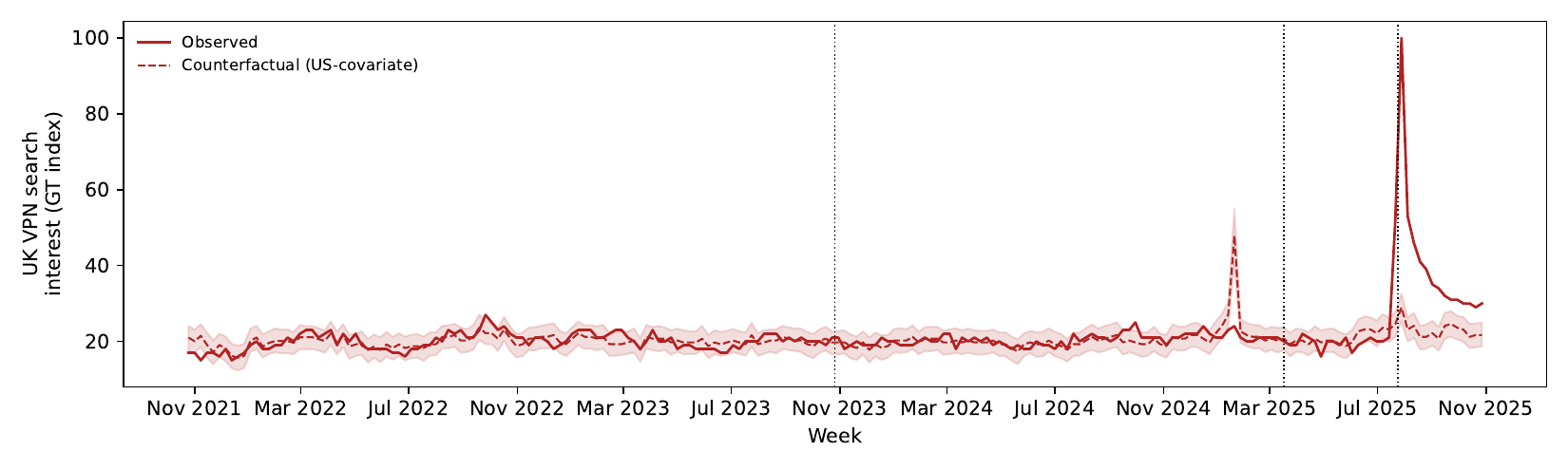}
\caption{UK VPN Google Trends search interest with US VPN search interest as covariate. Solid line: observed; dashed: BSTS counterfactual; shaded: 95\% credible interval; vertical lines mark OSA milestones.}
\label{fig:rq1_gt}
\end{figure}

\begin{table}[tb]
\centering
\caption{CausalImpact on UK VPN Google Trends search interest. (\S\ref{subsec:primary_spec}). $^{**}p<0.05$, $^{*}p<0.10$; ${\approx}$: not significant.}
\label{tab:trends}
\begin{tabular}{lrrr}
\toprule
\textbf{Specification} & \textbf{Oct 2023} & \textbf{Mar 2025} & \textbf{Jul 2025} \\
\midrule
No covariate      & ${\approx}+0\%$   & ${\approx}{-1\%}$ & $+147\%^{**}$ \\
US VPN covariate  & ${\approx}+5\%$   & ${\approx}+0\%$   & $+135\%^{**}$ \\
Lag-52 covariate  & ${\approx}{-2\%}$ & ${\approx}{-2\%}$ & $+144\%^{**}$ \\
\bottomrule
\end{tabular}
\end{table}

\para{Google Trends corroboration} The previous claims rely
on filtered Reddit data, with two possible weaknesses: Reddit's population bias and a classifier that might create the pattern. UK Google Trends search
interest (Table~\ref{tab:trends}) is less susceptible to either concern. It shows the same pattern: no
change at Royal Assent (${\approx}+0\%$) or Enforcement (${\approx}{-1\%}$), a
significant rise only at Age Verification ($+147\%$), holding across covariates
($+147\%$ none, $+135\%$ US search interest, $+144\%$ 52-week lag). Since the US
covariate absorbs global VPN demand yet barely moves the estimate, the rise is
UK-specific. An independent measure with entirely different biases reproducing
the same Age-Verification concentration is our strongest single evidence that the
finding is not an artifact of the measurement approach. (Figure~\ref{fig:rq1_gt}).
{Meanwhile, Google Trends' magnitude ($+147\%$) is far below the
classified, UK-authored VPN estimate ($+1265\%$), but it closely matches the
\emph{unfiltered} UK Reddit content effect ($+145\%$, Raw (UK); Table~\ref{tab:content_effects}). This may suggest the two unfiltered, population-level measures (Raw UK Reddit content and Google Trends) agree on the true magnitude, while the much larger classified+UK figure reflects two filters stacked together which leave a pre-period baseline of only a few documents per week (Table~\ref{tab:corpus_composition}). Percentage effects on that thin baseline are inherently more volatile than ones computed on a larger, unfiltered base.}

\subsection{Robustness}
\label{sec:rq1_d_robustness}
The headline Age Verification effect may be sensitive to {three} choices: the 8-week post-window, how much evidence we require before
calling an author UK-resident (i.e., how many posts and comments the author should have posted in UK geographical subreddits), {and the choice of observation-noise model in CausalImpact}. We varied each in turn, keeping the rest of the primary specification
fixed\textbf{; full matrices in
Appendices~\ref{appendix:extra_robustness} and~\ref{appendix:uk_author_robustness}.
None of them changes the conclusion.} \emph{Post-window:} the UK-authored VPN
effect is $+1132\%$, $+1265\%$, and $+932\%$ at 4, 8, and 12 weeks, and the
pattern (no effect at Royal Assent or Enforcement, a large effect at Age
Verification) holds at every width.
{
\emph{UK-residency threshold:} {$k$, the number of distinct UK geo-subreddits an author has posted in, is our confidence threshold for the UK-resident label. Higher $k$ indicates stronger evidence the author lives in the UK.} For VPN subreddits, increasing {k} raises the effect, from
+1256\% to +3188\% at \(k{\geq}5\)
(Table~\ref{tab:app_k_robustness}). This shows the effect is strongest among the authors we can most confidently identify as UK residents. The stronger effect is not an artefact of prolific posting: for VPN, mean author activity stays flat across $k$ even as the effect rises steadily. For Politics, author activity instead rises sharply with $k$ (Spearman $\rho{=}0.45$), yet the effect stays flat and non-monotonic ($+1450\%$ to $+1333\%$)  (Appendix~\ref{appendix:uk_author_robustness}).
{\emph{Observation-noise model:} the VPN-UK classified series is a low-count series (pre-period mean $\approx$7 documents/week), for which CausalImpact's default Gaussian observation model is an approximation. A variance-stabilizing $\log(1+x)$ transform, more appropriate for count data, yields a back-transformed relative effect of $+1399\%$, the same order of magnitude as the Gaussian estimate above (\S\ref{appendix:gaussian_robustness}).}

}

\section{RQ2: Discourse Framing and Positioning}
\label{subsec:framing_method}

    RQ2 asks how users frame the OSA, whether discussion is supportive or critical, and what arguments users advance. We conduct all RQ2 analyses on the UK-resident OSA-classified corpus constructed through the preprocessing pipeline above. This corpus contains 45,438 documents.

    We use Latent Dirichlet Allocation (LDA) to identify dominant themes in OSA-related discussion. LDA is applied separately to VPN and UK Politics subreddits because the same terms can encode different concerns across the two communities. Documents are lowercased and lemmatised prior to fitting, with English and domain-specific stopwords removed. We select the number of topics $K$ using $C_V$ coherence and report the selected topic structures in the RQ2 results section.

    To support qualitative interpretation, we identify representative documents for each topic using posterior topic probabilities. For each topic, we examine high-probability documents and use LLM-assisted summarisation to characterise the main arguments advanced by users. The LLM summaries are used to support interpretation of topic clusters; the substantive claims reported in the results are grounded in the underlying Reddit documents and representative examples.

   {The three subheadings under which RQ2 results are
reported are not the LLM's output categories. For each topic, the 15
highest-probability documents were summarised with a fixed prompt
asking for (1) the concerns and arguments users raise, (2) examples or
real-world cases they cite, and (3) the overall tone (prompt in
Appendix~\ref{appendix:prompts}). These per-topic summaries were then
grouped by the authors into three cross-cutting themes that recurred
across topics: objections to age verification, general OSA opinion,
and awareness of provider risk.}

    We classify sentiment using two complementary approaches. First, we apply \texttt{roberta-base-sentiment-latest}, a RoBERTa-base model fine-tuned on 124M tweets, producing three-class labels: negative, neutral, and positive. Second, we use Gemini~2.5~Flash to produce same three-class sentiment labels. Because Reddit comments are often difficult to interpret without context, each comment is prepended with its parent post title before sentiment classification when the parent title is available. Topic-level sentiment is computed by joining sentiment labels with LDA topic assignments on the UK-resident classified corpus.

\subsection{Results}
\label{sec:rq2}

We apply topic modelling and sentiment analysis to the
OSA-classified corpus from Section~\ref{sec:method} to characterise how users
frame the Act and why VPN { attention} is the dominant response.

\begin{figure}[t]
\centering
\includegraphics[width=\linewidth]{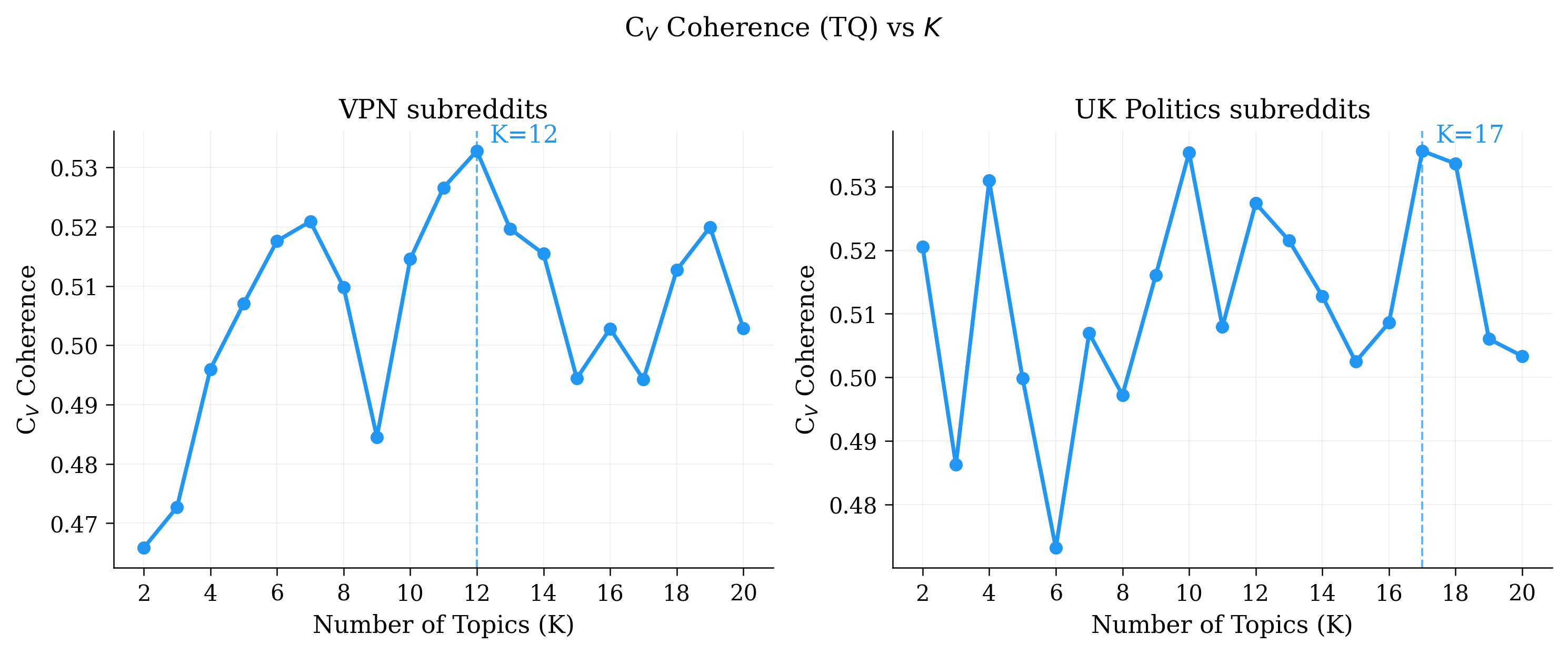}
\caption{C$_V$ coherence vs.\ number of topics $K$ for LDA,
across VPN and UK Politics subreddits.}
\label{fig:coherence}
\end{figure}

\para{Topic Modelling.}
\label{subsec:topics}
We apply LDA separately to VPN and UK Politics subreddits, selecting
$K$ using C$_V$ coherence over $K \in [2, 20]$ (Figure~\ref{fig:coherence}.)
On the selection corpus, coherence peaks at $K{=}12$ for VPN
(C$_V = 0.533$) and $K{=}17$ for Politics (C$_V = 0.536$).
{LDA topic assignment is performed on 44,311 of the
45,438 UK-resident documents; the remaining 1,127 documents are
dropped prior to fitting because they contain fewer than three tokens
after lemmatisation and stopword removal.}

Table~\ref{tab:topics_combined} shows the resulting topics. In VPN
subreddits, discussion is dominated by surveillance- and rights-oriented
framings: government surveillance \& OSA (19.4\%), surveillance \&
anonymity (19.1\%), age verification \& digital ID (9.8\%), and VPN
bans \& state censorship (5.0\%) together account for roughly 53\% of
VPN discussion. Provider-trust and encryption framings form a second
group (VPN no-logs \& court orders 9.2\%, end-to-end encryption 4.1\%),
and technical access and circumvention a third (VPN connectivity \&
blocking 8.6\%, ISP visibility/DNS/Tor 5.5\%, operational-security
practices 5.2\%, app/browser access 5.0\%, geo-blocking \& site access
5.0\%, censorship circumvention \& streaming 4.1\%). Notably, purely
access-oriented framings (geo-blocking \& site access, censorship
circumvention \& streaming) account for under 10\% combined, an order
of magnitude below the surveillance- and rights-oriented framings. The
dominant response is thus privacy-protective refusal rather than
access-seeking. In Politics subreddits the themes are legislative process
(OSA debate, Ofcom, parliamentary procedure), content regulation (age
verification, child safety, social media bans), and surveillance and
rights (facial recognition, government power, digital rights).

{To check that topic summaries (drawn from each cluster's
highest-probability documents) represent the wider cluster, we drew a
random sample of documents from every topic and manually judged
whether each matched the assigned theme. For the topics underlying our
qualitative findings, the random samples were consistently on-theme.
Two clusters were found to contain mostly non-substantive
content: the Parliamentary process cluster (1.3\%) consisted 
of an automated weekly digest of bills before Parliament, and
in the Technical circumvention cluster (3.5\%), 4 of the 10 sampled
documents were AutoModerator messages flagging duplicate posts rather
than user-written comments. Neither cluster contributes to the themes
reported in our qualitative findings.}

\begin{table}[h]
\centering
\small
\caption{LDA topics for VPN and UK Politics subreddits, grouped by cross-cutting theme, with prevalence.}
\label{tab:topics_combined}
\resizebox{\columnwidth}{!}{%
\begin{tabular}{ll@{\hspace{2pt}}r@{\hspace{2pt}}p{0.45\linewidth}r}
\toprule
\textbf{Subreddits} & \textbf{Theme} & \textbf{\%} & \textbf{Topic} & \textbf{\%} \\
\midrule
\multirow{12}{*}{\makecell[l]{VPN \\ subreddits}}
& \multirow{4}{*}{\makecell[l]{\footnotesize Surveillance \\ \footnotesize \& digital rights}} & \multirow{4}{*}{53.3}
  & Government surveillance \& OSA & 19.4 \\
& & & Surveillance \& anonymity & 19.1 \\
& & & Age verification \& digital ID & 9.8 \\
& & & VPN bans \& state censorship & 5.0 \\
\cmidrule(l){2-5}
& \multirow{2}{*}{\makecell[l]{\footnotesize Provider trust \\ \footnotesize \& encryption}} & \multirow{2}{*}{13.3}
  & VPN no-logs \& court orders & 9.2 \\
& & & End-to-end encryption & 4.1 \\
\cmidrule(l){2-5}
& \multirow{6}{*}{\makecell[l]{\footnotesize Technical access \\ \footnotesize \& circumvention}} & \multirow{6}{*}{33.4}
  & VPN connectivity \& blocking & 8.6 \\
& & & ISP visibility, DNS \& Tor & 5.5 \\
& & & Opsec \& anonymity practices & 5.2 \\
& & & App/browser access & 5.0 \\
& & & Geo-blocking \& site access & 5.0 \\
& & & Censorship circumvention \& streaming & 4.1 \\
\midrule
\multirow{17}{*}{\makecell[l]{UK Politics \\ subreddits}}
& \multirow{7}{*}{\makecell[l]{\footnotesize Legislative \& \\ \footnotesize regulatory process}} & \multirow{7}{*}{55.5}
  & OSA general discussion & 39.8 \\
& & & OSA parliamentary debate & 3.5 \\
& & & Platform compliance \& liability & 3.3 \\
& & & Ofcom \& broadcast regulation & 3.2 \\
& & & Ofcom notices \& media & 2.6 \\
& & & Government ministers \& offices & 1.8 \\
& & & Parliamentary process & 1.3 \\
\cmidrule(l){2-5}
& \multirow{4}{*}{\makecell[l]{\footnotesize Content regulation \\ \footnotesize \& child safety}} & \multirow{4}{*}{24.8}
  & Child safety \& age-restricted content & 11.9 \\
& & & Age verification mechanisms & 10.0 \\
& & & Online abuse \& harmful content & 1.6 \\
& & & Moderation & 1.3 \\
\cmidrule(l){2-5}
& \multirow{3}{*}{\makecell[l]{\footnotesize Circumvention \\ \footnotesize \& access}} & \multirow{3}{*}{11.5}
  & OSA circumvention \& petitions & 4.6 \\
& & & Technical circumvention & 3.5 \\
& & & Subscription costs \& ISP services & 3.4 \\
\cmidrule(l){2-5}
& \multirow{3}{*}{\makecell[l]{\footnotesize Surveillance}} & \multirow{3}{*}{8.2}
  & Free speech \& social media & 3.0 \\
& & & End-to-end encryption \& security & 2.8 \\
& & & Facial recognition \& policing & 2.4 \\
\bottomrule
\end{tabular}}
\end{table}

\para{Sentiment Analysis.}
\label{subsec:sentiment}
Sentiment is identified using two models applied to the 45,438 UK-resident documents: a RoBERTa-base model (\texttt{roberta-\linebreak[2]base-\linebreak[2]sentiment-\linebreak[2]latest}), and Gemini~2.5~Flash. The latter is prompted to assign
three-class labels (the full sentiment classification prompt is provided in Appendix~\ref{appendix:prompts}).

Table~\ref{tab:sentiment_overall} reports the results. Positive sentiment is negligible
under both models (4--7\%), with near-zero genuine pro-OSA voices, and negativity
dominates in both groups. The models diverge most on online abuse and harmful
content (RoBERTa 74\%, Gemini 9\%): posts there describe abuse and criminal
behaviour, so RoBERTa reads the negative wording as negative sentiment while
Gemini recognises the authors are discussing harms, not criticising the OSA, and labels most of the cluster neutral. Neutral content is substantial throughout and largely informational -- users explaining how age
verification works or what the Act covers rather than voicing an opinion -- while
the small positive minority chiefly supports child-protection aims even where the
same users criticise the Act's implementation; reporting both alongside the
negative share avoids overstating uniform opposition.

\begin{table}[h]
\centering
\footnotesize
\caption{Sentiment distribution by subreddit group and model (UK-resident authors).}
\label{tab:sentiment_overall}
\resizebox{\columnwidth}{!}{%
\begin{tabular}{llrrr}
\toprule
\textbf{Group} & \textbf{Model} & \textbf{Negative} & \textbf{Neutral} & \textbf{Positive} \\
\midrule
\multirow{2}{*}{VPN subreddits}         & RoBERTa & 49\% & 47\% & 4\% \\
                                         & Gemini  & 57\% & 38\% & 5\% \\
\midrule
\multirow{2}{*}{UK Politics subreddits} & RoBERTa & 55\% & 42\% & 4\% \\
                                         & Gemini  & 51\% & 42\% & 7\% \\
\bottomrule
\end{tabular}}
\end{table}

\begin{figure*}[t!]
\centering
\begin{subfigure}[t]{0.49\linewidth}
    \includegraphics[width=\linewidth]{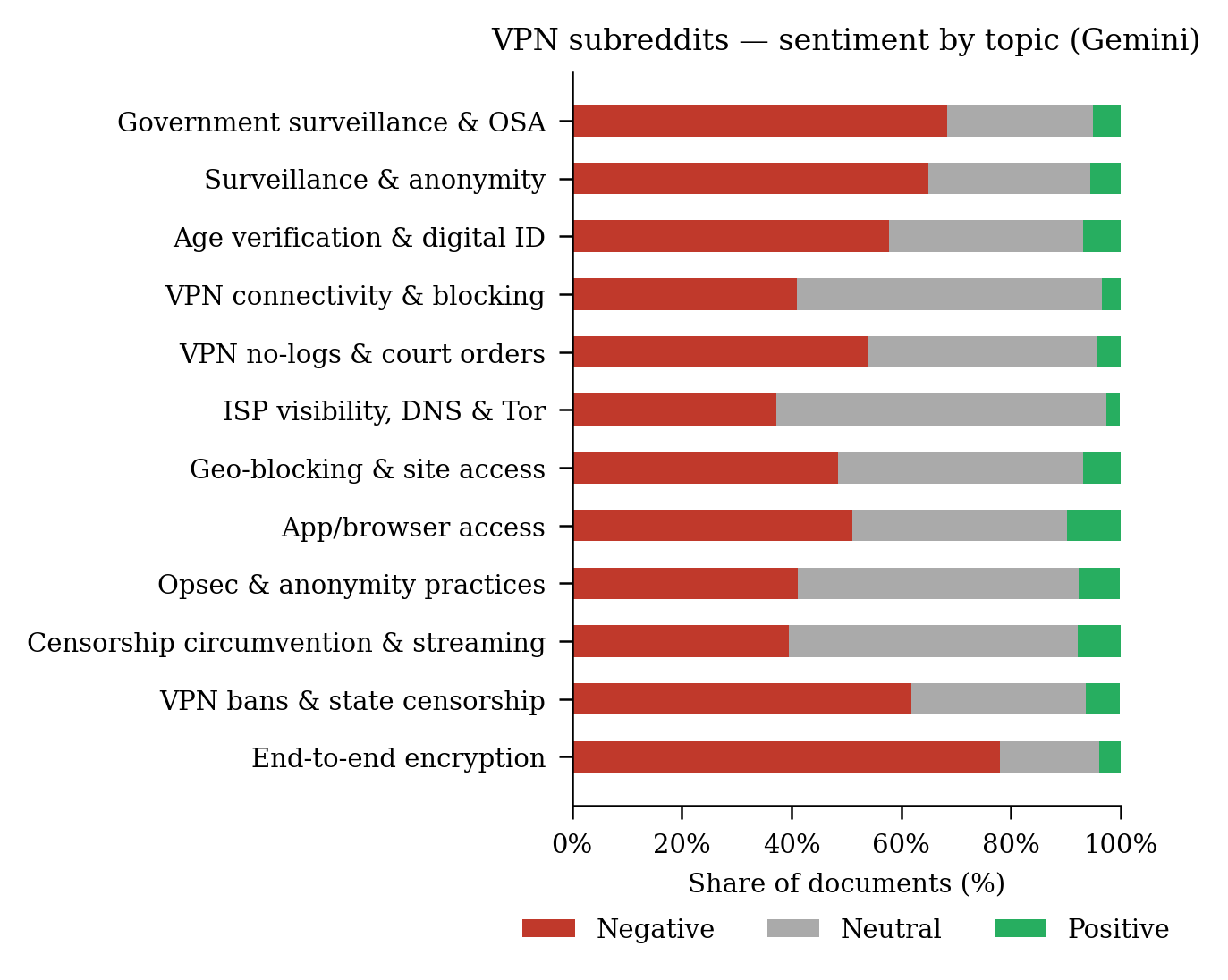}
    \caption{VPN subreddits.}
    \label{fig:topic_sentiment_vpn}
\end{subfigure}
\hfill
\begin{subfigure}[t]{0.49\linewidth}
    \includegraphics[width=\linewidth]{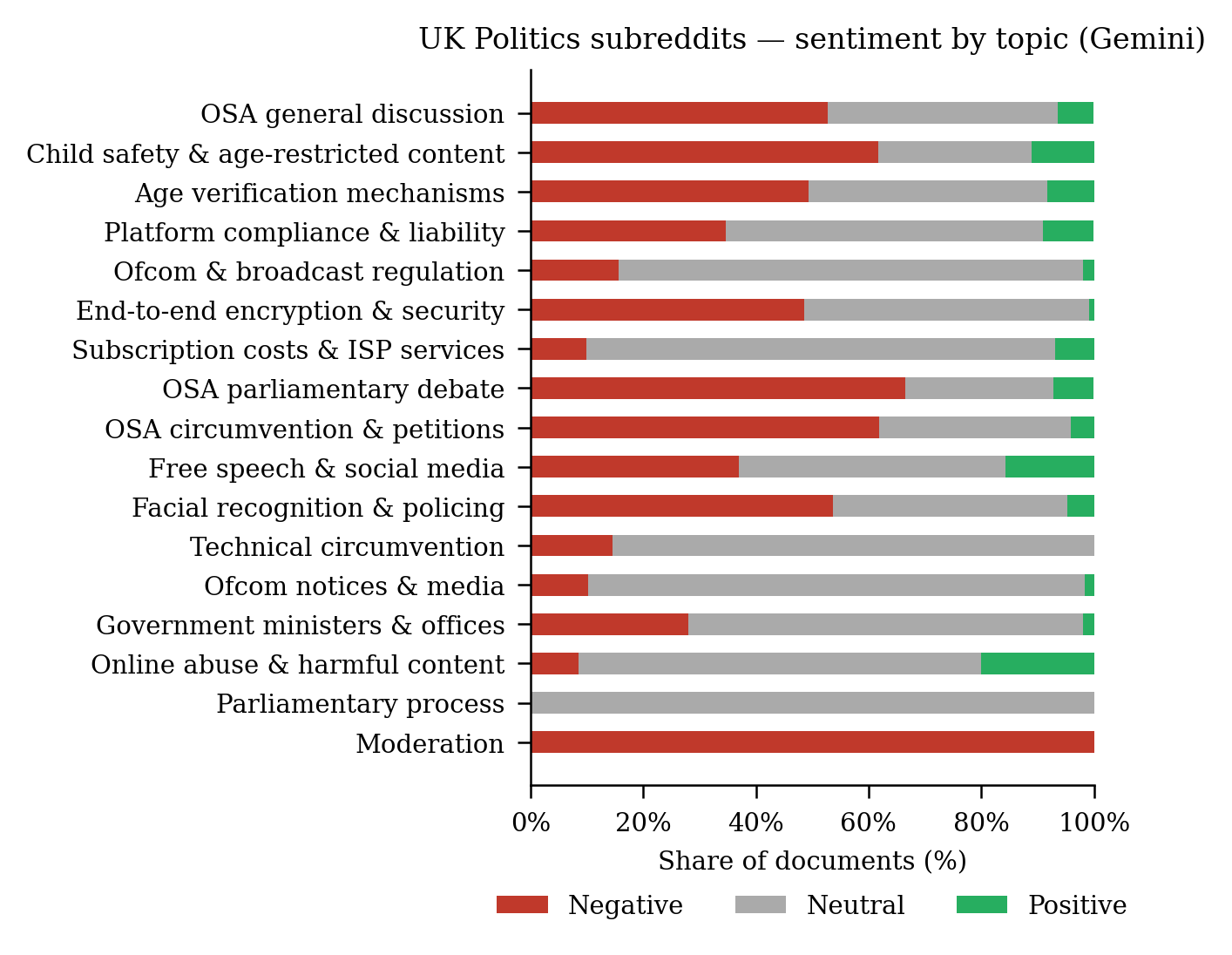}
    \caption{UK Politics subreddits.}
    \label{fig:topic_sentiment_politics}
\end{subfigure}
\caption{Sentiment by LDA topic for VPN and UK Politics subreddits, classified using Gemini.}
\label{fig:topic_sentiment_all}
\end{figure*}

Sentiment varies systematically across topics (Figure~\ref{fig:topic_sentiment_all}).
Both models agree that the two most prevalent VPN topics are also among the most
negatively framed: government surveillance \& OSA (RoBERTa
52\%, Gemini 68\%) and surveillance \& anonymity (RoBERTa 61\%, Gemini
65\%). Higher raw negativity appears in smaller, low-prevalence
clusters (VPN bans \& state censorship, end-to-end encryption). In
Politics, OSA parliamentary debate (RoBERTa 64\%, Gemini 67\%) and
child safety (RoBERTa 67\%, Gemini 62\%) rank highest.

{To contextualise our choice of sentiment model, we
compared the two classifiers across the 45,438 UK-resident documents.
The models agree on 63.7\% of documents (VPN 62.4\%, Politics 63.7\%),
leaving 16,547 documents (36.3\%) on which they diverge. To adjudicate
these disagreements, we drew a fresh random sample of 100 divergent
documents (50 VPN, 50 Politics) from the high-precision corpus for
blind human annotation.
The annotator agreed with Gemini in 63\% of cases and with RoBERTa in
31\% (6\% with neither), a preference consistent across both the
Politics (72\% vs.\ 24\%) and VPN (54\% vs.\ 38\%) subsets. This
approximately 2:1 preference supports our use of Gemini's context-aware
labels for the topic-level sentiment analysis in RQ2.}

\para{Why Do Users Prefer VPNs Over Facial Age Verification?}
\label{subsec:vpn_preference}{Our findings characterise how the OSA is discussed within VPN/privacy
and UK-politics communities, which are not neutral samples: users in
these spaces are more likely to frame VPN use around privacy and rights
rather than access to restricted content. Our within-cluster spot-check
confirms the corpus also contains straightforward access-oriented
VPN discussion (e.g. geo-unblocking). We
therefore report these as framings that are characteristic of these
communities, not as the distribution of motives among UK users at large.}
{One user argued the Act's real aim was population-wide
surveillance rather than child protection, doubting that any child of
the youngest generation would struggle to install a VPN. Another
rejected the premise that VPN use is mainly about pornography, framing
the Act as seizing on a single issue to undermine digital privacy
wholesale -- and arguing that if prohibition were the goal, the
government should legislate it directly.}
The age verification cluster reveals three distinct objections. First,
users see age verification as mandatory identification rather than a
proportionate check. Submitting a government ID or a selfie creates a
persistent record with a commercial intermediary that users feel unable
to retract. Second, facial recognition is treated as categorically more
invasive. Users argue they have a right not to be recorded at all,
not just a right not to be identified, and are concerned that AI
systems could scan faces without consent in public or semi-public
settings. Third, users question whether age verification targets the
right problem. A recurring argument in this cluster is that social-media
restrictions for minors would address children's exposure more directly
than age-gating adult sites, which users perceive as easy to bypass
with a free VPN.

The child safety cluster in Politics subreddits develops this further.
Users compare the OSA to age restrictions on alcohol and cigarettes.
These rules exist but do not stop determined teenagers. Blocking adult
sites, they argue, pushes children toward less regulated alternatives
where the risks are greater. The solution users propose is simpler:
mandatory ISP level filtering on by default, or splitting home routers
into adult and child networks. These would be harder to bypass and less
invasive for adults than submitting biometric data to a third party.
Several users also suggest the Act is less about protecting children
and more about creating infrastructure to track adult content
consumption. A VPN sidesteps all of this -- bypassing the age
verification block without any data exchange with Yoti, Veriff, or any
other intermediary -- which is why it becomes the default response to
the OSA rather than compliance. One
user warned that the Act would simply push people toward obscure,
less trustworthy sites in order to preserve access and anonymity --
pornography included, since the underlying objection is to the
verification mechanism, not to what it gates.

\para{What Do Users Think About the OSA?}
\label{subsec:osa_opinion}
Opposition to the Online Safety Act is consistent across both
communities. Users critique the legislative strategy itself rather than
defending access to restricted content.
{Users tended
not to defend pornography as such; one framed it as the canary in the
coal mine of censorship, warning that the same apparatus could later
be turned on books, films, or broadcasts the government disliked.}

The government surveillance cluster frames the OSA as part of a broader
anti-encryption push. The central concern is Section~122, which users
call the ``spy clause.'' It would require platforms to scan private
messages for illegal content using client-side scanning. This means
analysing messages on the device before or after encryption, which
experts and users alike argue is indistinguishable from breaking end to
end encryption entirely. WhatsApp and Signal both threatened to leave
the UK rather than comply, and the government paused enforcement of the
provision in September 2023, acknowledging the technology to do this
without compromising privacy does not yet exist. Users compare this to
the EU's Chat Control initiative and the US EARN IT Act, framing all
three as the same policy with different branding. The argument is not
that child safety does not matter but that scanning everyone's messages
is not a targeted intervention -- it is mass surveillance with a child
safety justification.

The parliamentary debate cluster focuses on the political failure.
Labour's position, that the OSA does not go far enough and that VPNs
should be banned, is cited as evidence that there is no mainstream
party to vote for if you oppose the Act. Ministers who accused OSA
critics of defending predators are treated as closing down the debate
rather than engaging with it. Some users say they will spoil their
ballot rather than vote for any party that supported the legislation.

The OSA general discussion cluster reflects a simpler frustration:
users do not think the government understands what it is regulating.
The Act is described as a political gesture rather than a technical
solution. The circumvention cluster adds that a petition to repeal the
OSA reached over 300,000 signatures. Users who use VPNs to get around
the Act do not frame it as breaking the rules. They frame it as a
rational response to a law they did not consent to and do not consider
legitimate.

\para{Are Users Aware of Risks from Higher-Risk VPNs?}
\label{subsec:vpn_risk_awareness}
Awareness of provider risk is present but unevenly distributed. The
no-logs and court-orders cluster contains technically informed discussion of
which VPN providers can actually be trusted not to hand over user data. Users express scepticism about providers that claim not to keep logs but
could still be compelled to begin doing so; they point to a major
privacy-focused provider that was ordered to log a specific user and hand
the data over. The variable users treat as decisive is jurisdiction --
where a provider's infrastructure sits and whether its legal framework
could compel disclosure -- rather than the assurances printed in a
privacy policy.
{Concern centres on free services rather than comparisons
between specific paid providers. One user warned that free app-store VPNs
may be insecure or even operated by a hostile state, noting that the people
age verification is meant to protect --- those without a means to pay --- are
precisely those most likely to install such apps. Another worried that free
VPNs would end up monetising children's data, and hoped first-time users would instead choose
privacy-focused options. Some free
providers make money by collecting and selling user data, which defeats the
purpose of using a VPN for privacy.}

The surveillance and anonymity cluster is more resigned. Users here argue that
meaningful anonymity was already gone before the OSA. Advertisers,
ISPs, and platforms already hold enough data to profile most people,
and the Act just adds another layer. Rather than expecting legislation
to fix this, these users are building their own workarounds. They use
pseudonymous email addresses, VPNs to reduce the data available to
advertisers, and open source tools that do not report back to a central
server. The gap between this group and users who are new to VPNs
because of the OSA is the core concern. Someone installing their first
VPN because an adult site now requires age verification has no reason to
know the difference between a provider with a verified no log audit and
a free app that sells browsing data. Whether the OSA's enforcement
milestones disproportionately drove users toward the latter is
addressed in Section~\ref{sec:rq3}. 

\section{{RQ3: Policy-Disclosed VPN Privacy Risk and Search Attention}}
\label{subsec:vpn_method}
\label{subsec:vpn_data}

    RQ3 asks whether the rise in VPN attention was concentrated among providers with weaker disclosed privacy practices. We answer this question in three steps. First, we analyse archived privacy-policy pages for 69 unique VPN services. Second, we classify each VPN service into a low, medium, or high disclosed privacy risk category using predefined privacy markers. Third, we compare monthly Google Trends attention across these risk categories, showing how attention changed across provider categories.

    \para{VPN privacy-policy records.} 
    In May 2026, we identified VPNs available in the UK through Google Play, the Apple App Store, Google Search, and ChatGPT recommendations. After deduplicating apps belonging to the same service, the corpus comprised 69 VPN services. For each service, we downloaded the privacy-policy HTML and recorded its URL, six extracted privacy markers with supporting evidence, disclosed privacy-risk category, and Google Trends query. Each VPN is represented by archived privacy-policy page, extracted privacy markers, assigned privacy-risk category, and a monthly UK Google Trends attention array. When multiple apps referred to the same service, we retained a single service-level entry for analysis. These records form the basis for analysing data collection practices, logging policies, tracking practices, retention rules, third-party sharing, and privacy guarantees.

    \para{Privacy-policy extraction and risk classification.} The goal of the trustworthiness analysis is to assess whether popular VPN services preserve user privacy or introduce additional privacy risks through their data collection, logging, tracking, retention, or sharing practices. We treat privacy policies as the provider's official disclosure of these practices and classify each VPN into one of three disclosed privacy-risk categories: Low Disclosed Risk, Medium Disclosed Risk, or High Disclosed Risk.
    
    We apply an automated two-stage analysis pipeline. In the first stage, privacy-policy text is processed using an LLM-based information-extraction pipeline. The extractor converts each policy into a structured set of privacy markers and returns short evidence snippets for the extracted labels for manual validation of LLM outputs after information extraction.
    In the second stage, these extracted markers are used in a rule-based privacy risk classification model to assign the final risk category.

    Table~\ref{tab:vpn_markers} summarises the privacy markers extracted from VPN privacy policies. For each marker, the model assigns one of three values: \textit{yes}, \textit{no}, or \textit{unclear}. A marker is labelled \textit{yes} only when the policy explicitly describes the practice. Ambiguous, vague, or contradictory statements are labelled \textit{unclear}. During risk scoring, unclear values are conservatively treated as absence of evidence rather than direct evidence of risk.

    \begin{table}[!tb]
        \centering
        \footnotesize
        \caption{Privacy markers from VPN privacy policies.}
        \label{tab:vpn_markers}
        \begin{tabular}{p{0.28\linewidth}p{0.64\linewidth}}
        \toprule
        \textbf{Marker} & \textbf{Description} \\
        \midrule
        \texttt{traffic\_logging} &
        VPN logs browsing activity, DNS queries, visited websites, or communication content. \\
        \addlinespace
        \texttt{connection\_metadata} &
        VPN collects connection metadata such as timestamps, session duration, bandwidth usage, or IP addresses. \\
        \addlinespace
        \texttt{tracking\_identifiers} &
        VPN collects persistent identifiers such as device IDs, advertising IDs, or fingerprints. \\
        \addlinespace
        \texttt{third\_party\_sharing} &
        VPN shares user data with advertisers, analytics providers, or other partners. \\
        \addlinespace
        \texttt{long\_term\_retention} &
        VPN retains user data or logs for extended or vaguely defined periods. \\
        \addlinespace
        \texttt{policy\_vagueness} &
        Policy contains contradictory, ambiguous, or misleading statements about logging or data usage. \\
        \bottomrule
        \end{tabular}
        \vspace{-10pt}
    \end{table}

    To improve extraction reliability, the model is instructed to distinguish between VPN service practices and general website practices. For example, website cookies, marketing analytics, or advertising trackers present on the provider's website are not interpreted as evidence of tracking by the VPN service itself unless the policy explicitly connects such practices to the VPN application or infrastructure. The extraction pipeline also requires the model to provide short verbatim evidence snippets supporting each extracted marker.

After extracting the relevant policy markers, we classify each VPN to a privacy-risk category using predefined rules. \textbf{High Disclosed Risk} applies where the policy explicitly states that traffic is logged, or where it indicates connection-metadata logging, long-term retention, and third-party sharing. \textbf{Medium Disclosed Risk} applies where the policy indicates connection-metadata logging, tracking identifiers, third-party sharing, or both vague policy language and long-term retention. \textbf{Low Disclosed Risk} applies where none of these practices is clearly disclosed.

    Formally, let $T$ denote traffic logging, $M$ connection metadata logging, $S$ third-party sharing, $R$ long-term retention, $I$ tracking identifiers, and $V$ policy vagueness. Each variable takes the value 1 if the corresponding marker is present and 0 otherwise. The privacy risk classification function is defined as:

        \[
        \text{Risk}(\mathrm{VPN}) =
        \begin{cases}
        \text{High}, & T \lor (M \land R \land S) \\
        \text{Medium}, & M \lor I \lor S \lor (V \land R) \\
        \text{Low}, & \text{otherwise}
        \end{cases}
        \]

    \para{Provider popularity.} 
    After assigning risk categories, we analyse how search attention to VPN providers changed over time. We use saved Google Trends data collected through pytrends. For each provider name, we use monthly UK search interest and link the resulting series to the provider’s risk category. Unlike RQ1 and RQ2, which are bounded by Reddit corpus collection, RQ3's outcome measure is Google Trends search interest, a data source with no comparable collection cutoff. We therefore extend the RQ3 analysis window through July 2026, rather than truncating it to match the October 2025 Reddit corpus boundary. The extended window lets us observe whether VPN search attention shows a corresponding movement around the same period, using an independent, differently-bounded data source.

    Google Trends reports relative search interest rather than absolute search volume. In addition, provider terms were queried in batches, and Google Trends normalises each batch separately. For this reason, the RQ3 analysis should be read as descriptive evidence of changing attention, not as a precise measure of installations, subscriptions, market share, or provider-level causal effects.

\subsection{Results}
\label{sec:rq3}

Increased VPN attention driven by regulation may not be distributed evenly across providers. Users who are new to VPNs may rely on app-store rankings, advertising, or search visibility rather than technical indicators of trustworthiness, making them more susceptible to providers with weak privacy practices. We therefore examine whether OSA-related increases in VPN attention were concentrated among higher-risk providers or instead reflected a broader increase in VPN demand across the market.

The analysed RQ3 dataset contains 69 archived VPN policy records (see Table \ref{tab:vpn_risk_categories}). Applying the rule-based classifier produced 26 Low Disclosed Risk VPN services, 35 Medium Disclosed Risk VPN services, and 8 High Disclosed Risk VPNs.

The most common positive risk markers were third-party sharing, which appeared in 30 records. Connection metadata logging, which appeared in 29 records. Tracking identifiers appeared in 24 records. Traffic logging was less common, appearing in 4 records. Long-term retention and policy vagueness appeared for 11 VPN services each. Many privacy-policy pages were unclear on at least one marker, especially retention and tracking. In the scoring rules, unclear values were treated as absence of positive evidence rather than direct evidence of risk.

\begin{table}[t]
\centering
\small
\caption{{Disclosed} privacy-risk categories in VPN dataset.}
\label{tab:vpn_risk_categories}
\begin{tabular}{l r p{0.48\columnwidth}}
\toprule
Risk category & VPNs & Meaning \\
\midrule
Low Disclosed Risk & 26 & No risk marker is clearly indicated. \\
Medium Disclosed Risk & 35 & At least one medium-risk marker is clearly indicated. \\
High Disclosed Risk & 8 & Traffic logging is indicated, or metadata logging, retention, and third-party sharing are all indicated. \\
\bottomrule
\end{tabular}
\end{table}

To evaluate extraction quality, we randomly selected six of the 69 provider-policy records. For each record, we compared all six extracted privacy markers and their supporting evidence snippets with the original archived policy. The policy evidence supported the markers in all records, and we found no extraction errors.

\begin{figure*}[htb!]
    \centering
    \includegraphics[width=\textwidth]{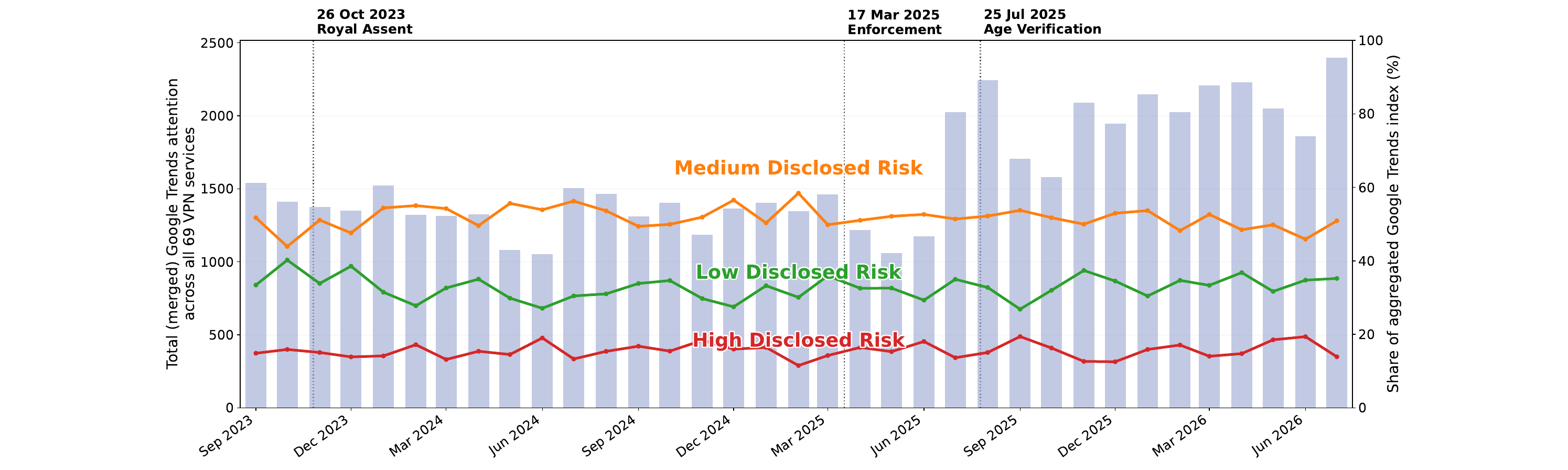}
    \caption{Monthly Google Trends attention to VPN services, September 2023--July 2026. Bars (left axis): total aggregated attention across all 69 VPNs. Lines (right): each disclosed-risk category's share of the total. Vertical lines are the OSA milestones.}
    \label{fig:trends}
\end{figure*}

We plotted extracted Google Trends linked with the privacy risks on Figure~\ref{fig:trends}, which shows both the aggregated Google Trends attention across all 69 VPN services and the proportional share of that attention accounted for by each disclosed-risk category over time. Total attention to VPN services rises substantially following the Age Verification deadline and remains durably elevated through the end of the observation window, rather than reverting to the pre-milestone baseline. This elevated attention persists through the most recent available data (July 2026), with no sign of decaying back toward pre-Age-Verification levels. {This sustained, undiminished elevation instead indicates a durable shift in attention to VPN services rather than a temporary reaction to coverage of the deadline itself.} The share of attention accounted for by low-, medium-, and high-disclosed-risk providers remains broadly stable. We therefore find no evidence that attention shifted disproportionately toward providers with higher disclosed privacy risk, even as overall attention to VPN services increased.

\label{subsec:vpn_trustworthiness}

\section{Discussion and Policy Recommendations}
\label{sec:discussion}

    {By linking changes in VPN search attention around regulatory milestones with providers’ privacy-policy disclosures, our approach helps the PETS community study the privacy risks that may result from using VPNs to bypass restrictions. The most immediate
implication is a corrective one: regulators and press coverage tend to
read VPN spikes as access-seeking, whereas the framing we recover is
privacy-protective. The displacement complicates how
VPN uptake is usually modelled. Prior work frames it as a deliberate
choice in which users weigh a tool's guarantees against cost and
trust~\cite{xue2024bridging, hobbs2018sudden}, whereas regulation produces users
pushed toward a VPN to avoid an identity check. The property
experienced users treat as load-bearing, jurisdiction, is precisely
what a first-time installer cannot read off an app-store listing.
The information asymmetry is therefore widest for exactly the users
regulation delivers, who reach these tools through app-store rankings
rather than the audit reports and no-logs disclosures that would let
them tell providers apart.} Taken together, the results show that the OSA generated measurable behavioural and discursive displacement. Our findings suggest that online safety regulation should be evaluated not only by platform compliance, but also by foreseeable user adaptation and the privacy risks created when users move to circumvention intermediaries.

\label{sec:rq4}
    We provide policy recommendations for policymakers, civil society, and the VPN industry, based on the displacement effects identified in our findings. More specifically, our results suggest that age-assurance regulation should be evaluated not only by formal platform compliance, but also by its behavioural first- or second-order effects. Where users respond to age checks through circumvention or migration to alternative intermediaries, the regulatory intervention may generate secondary privacy risks. Assessing effectiveness requires attention to likely user responses, including evasion, displacement, and reliance on less accountable services.

\para{For Policymakers and Regulators.}
\label{subsec:policy_legislators}
    Assessing effectiveness against circumvention is no longer a novel proposal: under the OSA, regulated services must already consider circumvention as part of their Children's Access Assessments, and Ofcom has committed to reviewing the effectiveness of age assurance and the factors hindering it. The open question is not \textit{whether} circumvention should be assessed but \textit{how} the displacement we document should bear on the next regulatory step. That step is now concrete. The UK government's 2026 consultation on children's online safety expressly asks whether VPNs should be age-restricted and what the privacy implications of such a requirement would be for the adult population [6, 26]. Our findings speak directly to that question. We show that the OSA's enforcement -- and most sharply its July 2025 age-verification deadline -- produced a large, concentrated increase in VPN attention (RQ1), that this attention is driven by privacy and anti-surveillance concerns rather than a wish to access restricted content (RQ2), and that the resulting {demand did not shift toward higher-risk VPN providers} but raised demand across the board (RQ3). The policy implication is that extending age assurance to VPNs would impose an identity- or biometric-verification burden on the entire adult VPN-using population {to} address a child circumvention problem, while reproducing at scale the very privacy externality the original measure generated. Regulators should therefore treat circumvention not as a residual enforcement gap to be closed, but as evidence that the chosen mechanism, i.e. geo-located age assurance discharged through commercial intermediaries, carries privacy costs that any extension to circumvention tools would compound.
    
    This issue is especially important when regulators consider extending age-verification obligations to VPNs. Although the UK government has consulted on age verification for VPNs \cite{UKParliamentChildrenWellbeingSchoolsLordsAmendments2026, DSITGrowingUpOnlineWorld2026}, this approach may trigger a technical arms race. VPN traffic can be obfuscated, services may change endpoints {or} ports, and users may shift to other proxy or tunnelling technologies. The key question is not only whether access controls can be imposed, but whether they remain effective given foreseeable circumvention.
    
    Credential sharing presents another limitation. Users may share passwords or authentication credentials, including payment card details, allowing unauthorised or underage individuals to bypass access controls. Policymakers should not treat successful authentication as proof that the user is the intended account holder or that the user meets the age requirement. Age-assurance frameworks should {treat} credential reuse and shared access as foreseeable sources of error when evaluating system effectiveness.
    
    These displacement effects have broader policy implications. Since RQ1 shows that major OSA milestones increased VPN-related attention, regulators should assess age-assurance systems not only for platform compliance but also for their effectiveness when users can circumvent them via VPNs, proxies, or other tunnelling tools. In post-implementation reviews, Ofcom should examine whether age-verification duties unintentionally drive users toward circumvention services that undermine child-safety protections and regulatory oversight.

\para{For Civil Society.}
\label{subsec:policy_civil_society}
    RQ2 indicates that while many users understand VPN risks, those prompted by OSA age verification may have limited technical knowledge and digital literacy. Civil society should intervene when these users are displaced, rather than relying solely on general digital literacy campaigns. Guidance should clarify VPN limitations, explain how free VPNs may monetise user data, and highlight the importance of audits, ownership transparency, and retention policies.
    
    Any extension of age-assurance obligations to circumvention technologies should carry proportionate transparency requirements, but the obligation must attach to \textit{availability in the UK market} rather than UK establishment. This distinction is decisive. Our RQ2 findings show that the variable users correctly treat as load-bearing is jurisdiction: no-log claims hold only so far as local law permits, and providers based in Five or Fourteen Eyes states are treated as categorically riskier regardless of policy text. The opaque free providers that constitute the actual privacy hazard are, by design, incorporated offshore precisely to sit beyond disclosure-compulsion regimes. A duty pegged to UK incorporation would therefore reach the reputable, already-audited providers while missing exactly the tier that prompts concern. The OSA already supplies the appropriate jurisdictional basis: a service is in scope where it targets UK users, has a significant UK user base, or is accessible in the UK and presents a material risk of harm. A transparency duty on VPN providers should track the same test, applying to any provider offering its service to UK users (including through listing in the UK Apple App Store or Google Play Store) irrespective of where it is incorporated. Because the offshore free applications that drive our privacy concern reach UK users predominantly through those stores (indeed, our own provider corpus in RQ3 was constructed in part from app-store rankings), the stores are the practical enforcement point: providers that fail to disclose logging practices, retention periods, parent-company ownership, and applicable jurisdiction could be made ineligible for UK distribution. Disclosure obligations framed this way would attach to market access rather than to a corporate footprint that the riskiest providers deliberately avoid.
    
 We do not overstate the reach of this mechanism. Providers distributed outside app stores, through direct download or sideloading, remain harder to compel, and the most determined opaque operators may route around any distribution channel. A market-availability duty is therefore necessary but not sufficient: it closes the dominant acquisition path for the less technically literate users our findings identify as most exposed, while leaving a residual margin that only international coordination could address.

\para{For VPN Providers.}
Reputable VPN providers should use the OSA enforcement period to differentiate themselves from opaque and unaudited providers that lead the app store rankings. We recommend three practical measures. 
First, providers should commission regular independent security and no-logs audits from recognised firms and publish the results, including the audit scope, methodology, and any vulnerabilities found before remediation. Providers should also issue annual transparency reports that disclose the number of law-enforcement or regulatory data requests received, the legal basis for any requests granted, and whether any user data was disclosed. Several providers, such as IVPN, ExpressVPN, and CyberGhost, already publish these reports. This practice should be a baseline expectation, not a market differentiator.
Second, providers should make their privacy policies significantly clearer.
Users have a complex relationship with data collection and especially value
the protection of sensitive personal data
\cite{kalameyets2026playing}. {Prior work shows that policy format affects how
efficiently and accurately readers can locate relevant information
\cite{mcdonald2009comparative}; studies of online advertising policies also
show that published policies may leave important data practices difficult to
determine \cite{cranor2014worth}.} Privacy policies should clearly state what
data is collected, how long it is retained, whether it is shared with third
parties, and the legal circumstances under which it may be disclosed to public
authorities.
Third, providers should participate in standardised certification schemes. The Internet Infrastructure Coalition's VPN Trust Initiative (VTI) provides an industry-led framework covering security, privacy, advertising practices, disclosure and transparency, and social responsibility~\cite{i2coalition2022vti}. Its VPN Trust Seal offers visible accreditation for providers that meet baseline standards and are subject to ongoing oversight, including unannounced audits. In the OSA context, such certification is especially important because increased demand for VPN services may lead users to unaccredited providers. Broader adoption of the VTI Principles would reduce the information asymmetry that makes it difficult for users to distinguish trustworthy services from opaque or exploitative ones.

\section{Limitations}
\label{subsec:limitations}
Reddit users skew younger, more technically literate, and more English-speaking than the UK population~\cite{proferes2021studyingreddit}, so our findings may not generalise to the broader public. The classifier validation should be interpreted as indicative rather than a gold-standard evaluation. The UK-resident filter is an imperfect proxy: users who post in UK geographic subreddits are likely but not guaranteed to be UK residents, and UK users who never post in geographic subreddits are missed, so both false positives and false negatives are present. The RQ3 popularity analysis relies on Google Trends, which measures relative search interest rather than installations, subscriptions, or active VPN use, and therefore captures attention to providers, not adoption. Lastly, the provider-risk classification is based on public privacy policies and disclosed ownership or audit information; they may not fully reflect logging, retention, or sharing practices, and opaque providers may under-disclose practices.

\section{Conclusion}
\label{sec:conclusion}

The UK Online Safety Act is a landmark intervention in platform regulation and age-assurance policy. This paper examined whether such regulation can produce secondary privacy risks by changing user behaviour. Using Reddit data, UK Google Trends search interest, Bayesian Structural Time Series models, topic modelling, sentiment analysis, and VPN privacy-policy classification, we studied behavioural and discursive responses across three OSA milestones.

We find that OSA enforcement produced a significant increase in VPN and privacy
discussion that was concentrated at the age-verification deadline. The discourse is predominantly critical of the Act: users frame VPN use as a privacy-preserving response to identity checks, facial age verification, surveillance, and perceived censorship. Our provider-level analysis shows that VPN attention increased across risk categories, but does not provide clear evidence of a systematic proportional shift toward higher-risk providers. Even so, the overall increase in VPN demand matters because users displaced toward commercial privacy tools may not understand the differences between audited no-log providers and opaque services with weaker privacy guarantees.

These findings show that online safety regulation should be evaluated beyond formal platform compliance. Where access controls lead users to circumvention tools, regulators must account for displacement, evasion, and the privacy risks created by reliance on new intermediaries. Future work should track VPN attention beyond the present data window, combine platform traces with interviews of UK VPN users, examine actual VPN installation or subscription data where available, and compare the UK case with similar regulatory interventions in other jurisdictions.

\begin{ethics}
\label{subsec:ethics}

All data used in this study consists of publicly accessible Reddit
posts and comments. We did not access private messages, use
authentication to retrieve restricted content, or interact with any
users. Reddit's Privacy Policy explicitly permits public
contributions to be shared and analysed.

No individual users are identified or reported on by name. Usernames
are retained internally for deduplication and the UK geo-filter but are
not published. All user opinions are
paraphrased to prevent search engine re-identification~\cite{fiesler2024remember}.
Only aggregate results are reported.

This study does not involve direct interaction with human participants
and does not meet the standard regulatory definition of human subjects
research~\cite{fiesler2024remember}. No IRB review was sought on this
basis, consistent with established practice in computational social
science research using public platform data.

\end{ethics}

\begin{openscience}
Upon acceptance, we will release the following research artifacts with all usernames removed: the LLM classification outputs for VPN and UK Politics subreddits; the UK-resident author corpus used for RQ2, along with sentiment labels, LDA topic assignments and all code used to derive them; RQ3 dataset with privacy polices, privacy markers and VPNs risk levels. To minimise re‑identification risk, our released artifacts contain only document identifiers, derived labels, and subreddits — not post or comment text, timestamps, or author identifiers — and we do not publish any UK‑resident author list; only aggregate results are reported in the paper.

\end{openscience}

\begin{ai}

    AI-based tools were used in both the research workflow and the writing process. 
    
    For research, we used LLM-based tools to support Reddit relevance classification, topic-cluster summarisation, sentiment classification, and VPN privacy-policy marker extraction. Sentiment analysis used a RoBERTa-based model and Gemini. The RQ3 privacy-policy extraction pipeline used GPT-5.5. OpenAI Codex with GPT-5.5 was used for coding.
    
    For writing, ChatGPT was used as an assistant for grammar and clarity. All edits were reviewed and accepted by the authors, who take full responsibility for the final content.
\end{ai}

\bibliographystyle{ACM-Reference-Format}
\bibliography{references}

\appendix
\section*{Appendix}

\section{Additional Robustness Analyses}
\label{appendix:extra_robustness}

This appendix collects additional robustness analyses supporting the RQ1
results. Unless noted otherwise, all checks use the primary specification
(no covariate, growing pre-period, 8-week week-aligned post-window).

\begin{figure*}[t]
\centering
\includegraphics[width=0.85\textwidth]{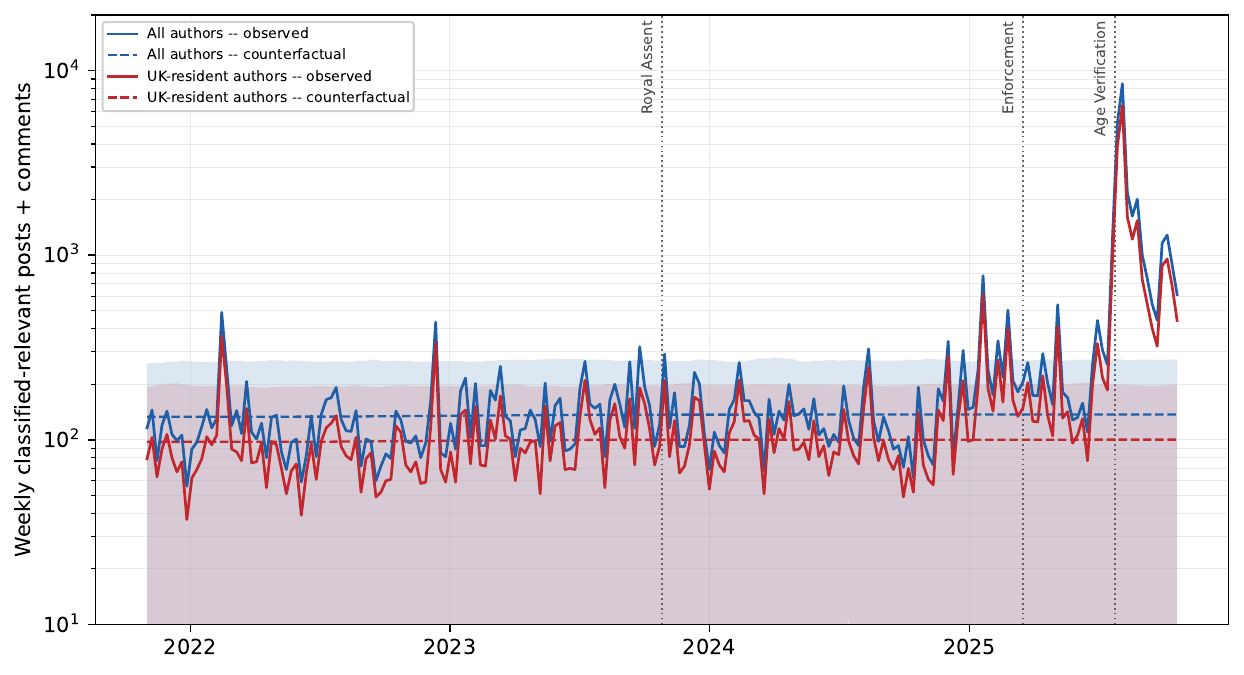}
\caption{OSA-classified content: CausalImpact timeseries for UK Politics subreddits (companion to Figure~\ref{fig:rq1_vpn_classified} in the main text). Blue: all authors; red: UK-resident authors. Solid lines show observed volume, dashed lines the BSTS counterfactual; vertical lines mark OSA milestones. $y$-axis is logarithmic (floored at 10) to show both the small pre-period baseline and the Age Verification spike on one scale. Shaded regions show the 95\% credible interval smoothed with a 5-week rolling window for legibility and clipped at the axis floor; the underlying weekly interval is wide and noisy relative to the near-zero baseline. Same design as Figure~\ref{fig:rq1_vpn_classified} -- a single continuous BSTS fit with one pre-period ending at Royal Assent and one post-period extending to the end of the study window -- and, for the same reason, not directly comparable, milestone-by-milestone, to the tabulated per-milestone percentages in Table~\ref{tab:content_effects}.}
\label{fig:rq1_politics_classified_companion}
\end{figure*}

\subsection{\rev{Royal Assent and Enforcement: Why the Positive Estimates Are Not OSA Effects}}
\label{appendix:royal_assent_null}

\rev{Several rows show significant positive effects at Royal Assent: keyword-matched
posts in both groups ($+20\%$ VPN, $+21\%$ Politics) and Politics user counts
($+24\%$ all authors, $+28\%$ UK). This appendix documents why we treat these as
artefacts of the specification rather than as regulatory effects. The argument does
not depend on judgement: both diagnostics below are mechanical.}

\rev{\para{The Politics estimate is not an increase.} Keyword-matched Politics posts
averaged $99.2$ per week over the twelve weeks before Royal Assent and $87.1$ per
week over the eight weeks after -- a $12\%$ \emph{fall}. The reported effect is
positive only because the fitted counterfactual ($72.6$ per week) sits near the
two-year pre-period mean ($70.7$) rather than near the immediately preceding level.
The series had stepped up during the Bill's final parliamentary passage: quarterly
mean weekly matched posts ran $61.8$, $64.9$, $71.2$, $62.0$, $75.5$ and $64.8$ from
Q1~2022 through Q2~2023, then jumped to $98.8$ in Q3~2023 -- entirely before the
26~October milestone -- and stayed near that level afterwards. With a growing
pre-period spanning two years, a level shift confined to the final quarter is
outvoted by the preceding twenty-one months, so the residual elevation is attributed
to the milestone that happens to follow it.}

\rev{\para{Both estimates fail a placebo-date test.} Re-estimating at quarterly dates
where no OSA event occurred (Table~\ref{tab:app_placebo_ra}) produces effects that are
frequently as large as, or larger than, the Royal Assent estimate itself. For Politics
the placebo three months earlier ($+42.6\%$) and the one three months later ($+47.6\%$)
both exceed Royal Assent's $+20.4\%$; for VPN the placebo three months later
($+24.9\%$) exceeds it as well. A milestone whose estimated effect is bracketed by
larger effects at adjacent non-events is not identified by this design.}

\rev{\para{Enforcement fails the same test.} The March 2025 illegal-harms deadline
behaves identically (Table~\ref{tab:app_placebo_ra}b). Its Politics estimate
($+65.7\%$ all authors, $+60.3\%$ UK) is comfortably exceeded by a placebo three
months earlier ($+123.0\%$ and $+128.8\%$), and another placebo returns a
\emph{significant negative} ($-27.5\%$). On the VPN side the milestone itself is null
($-12.4\%$) while two non-events return significant positives ($+41.0\%$,
$+47.7\%$). We therefore do not interpret the Enforcement estimates as regulatory
effects either.}

\rev{\para{Why this happens.} Both series trend upward across the study window: VPN
keyword-matched posts rise from a quarterly mean of $220$ per week in late 2021 to
$268$ in 2022, $285$ in Q3~2023 and $345$ in 2024. A counterfactual anchored on a
long, growing pre-period systematically under-predicts a trending series, so any
8-week window placed in a rising stretch returns a positive, often significant,
effect. This is a property of the pre-period design rather than of the data, and it
argues for reading the milestone estimates against the placebo column rather than
against zero.}

\rev{\para{Scope.} This does not implicate Age Verification, which for VPN is the one milestone of the three that passes this test. VPN's placebo dates are
null at ten and twelve weeks prior and only marginally significant at eight
($+41\%$, $p{=}0.083$; \S\ref{appendix:age_verification_manipulation_check}), against
a post-milestone effect an order of magnitude larger; the rise there is concentrated
in OSA-relevant content rather than spread across the subreddit group; and it is
corroborated by an independent, classifier-free outcome (Table~\ref{tab:trends}).
Politics does not pass this specific check as cleanly: its placebo eight weeks prior is
significant and sizeable ($+125\%$, $p{=}0.001$; Table~\ref{tab:app_placebo_ra}c). We
still read the Age Verification estimate as substantively real for Politics -- the
milestone effect remains an order of magnitude larger, and \S\ref{appendix:age_verification_manipulation_check}
gives independent, content-level evidence for it -- but, unlike VPN, we cannot claim
a clean pre-trend on this test alone.}

\begin{table}[tb]
\centering
\caption{\rev{Placebo-date and date-sensitivity test, 8-week window, UK-authored classified volume. Each non-bold row re-estimates the model at a date where no OSA milestone occurred. Panel (a) uses keyword-matched posts around Royal Assent; panel (b) uses classified-relevant content (total) around Enforcement; panel (c) uses UK-authored classified volume around Age Verification. In panels (a) and (b) the effects at non-events are comparable to or larger than the milestone estimate itself. Panel (c) shows this pattern too for VPN, where the closest placebo (8 weeks prior) is only marginally significant and over an order of magnitude smaller than the milestone effect; Politics instead shows a significant pre-trend at 8 weeks prior ($+125\%$, $p{=}0.001$), so this specific check does not clear Politics as a clean intervention the way it does VPN. Panel (c)'s four bold rows also serve as a date-sensitivity check: they re-estimate the milestone effect using each platform's reported Age Verification compliance date as the treatment date instead of a single fixed date, and are stable within each group (VPN $+1248\%$ to $+1261\%$; Politics $+1470\%$ to $+1490\%$) since all four dates fall in the same weekly bin. $^{**}p<0.05$, $^{*}p<0.10$.}}
\label{tab:app_placebo_ra}
\begin{tabular}{lrr}
\toprule
\textbf{Treatment date} & \textbf{VPN} & \textbf{Politics} \\
\midrule
\multicolumn{3}{l}{\textit{(a) around Royal Assent -- keyword-matched posts}} \\
25 Apr 2022 (placebo)  & $+0.7\%$  & $+13.6\%^{**}$ \\
25 Jul 2022 (placebo)  & $+5.4\%$  & $+11.5\%$ \\
24 Oct 2022 (placebo)  & $+2.9\%$  & $+1.2\%$ \\
23 Jan 2023 (placebo)  & $+4.0\%$  & $+14.8\%^{*}$ \\
24 Apr 2023 (placebo)  & $-0.8\%$  & $-9.0\%$ \\
24 Jul 2023 (placebo)  & $+11.0\%^{**}$ & $+42.6\%^{**}$ \\
\midrule
\textbf{26 Oct 2023 (Royal Assent)} & \textbf{$+20.4\%^{**}$} & \textbf{$+20.4\%^{**}$} \\
\midrule
22 Jan 2024 (placebo)  & $+24.9\%^{**}$ & $+47.6\%^{**}$ \\
22 Apr 2024 (placebo)  & $+4.1\%$  & $+19.6\%^{**}$ \\
\midrule
\multicolumn{3}{l}{\textit{(b) around Enforcement -- classified-relevant, total}} \\
18 Mar 2024 (placebo)  & $+13.6\%$ & $+9.3\%$ \\
17 Jun 2024 (placebo)  & $+41.0\%^{**}$ & $+14.7\%$ \\
16 Sep 2024 (placebo)  & $+17.2\%^{*}$ & $-27.5\%^{**}$ \\
16 Dec 2024 (placebo)  & $+47.7\%^{**}$ & $+123.0\%^{**}$ \\
\textbf{17 Mar 2025 (Enforcement)} & \textbf{$-12.4\%$} & \textbf{$+65.7\%^{**}$} \\
5 May 2025 (placebo)   & $-4.3\%$  & $+23.5\%$ \\
\midrule
\multicolumn{3}{l}{\textit{(c) around Age Verification -- UK-authored classified volume}} \\
1 May 2025 (placebo)   & $+11\%$  & $+38\%^{*}$ \\
15 May 2025 (placebo)  & $+27\%$  & $+36\%^{*}$ \\
29 May 2025 (placebo)  & $+41\%^{*}$ & $+125\%^{**}$ \\
\midrule
\textbf{Discord (21 Jul)} & \textbf{$+1248\%^{**}$} & \textbf{$+1476\%^{**}$} \\
\textbf{Aylo/Pornhub (24 Jul)} & \textbf{$+1261\%^{**}$} & \textbf{$+1470\%^{**}$} \\
\textbf{Ofcom deadline / Reddit (25 Jul)} & \textbf{$+1252\%^{**}$} & \textbf{$+1490\%^{**}$} \\
\textbf{X (26 Jul)} & \textbf{$+1261\%^{**}$} & \textbf{$+1474\%^{**}$} \\
\bottomrule
\end{tabular}
\end{table}

\para{What actually happened: r/privacy drives the raw pre-milestone declines.} A single subreddit, r/privacy, accounts for the largest share of VPN-relevant volume throughout the study period, and inspecting it directly explains the raw declines above without invoking the model. Before Royal Assent, r/privacy's moderators posted publicly on 13 October 2023 (516 upvotes) that the subreddit was switching to links-only submissions for the remainder of the month because of a moderator-capacity shortage, and confirmed the reversal on 2 November -- the exact week volume recovers. This is not a collection artefact: r/privacy's October 2023 raw counts (420 submissions, 6{,}178 comments) match a live re-query of the Arctic Shift API (420 submissions, 6{,}179 comments) to within one post. Before Enforcement, r/privacy again drives most of the decline, but the mechanism differs: rather than one dateable policy change, several unrelated general privacy/surveillance stories went unusually viral in January 2025 -- a post on AI-driven behavioural surveillance reached $8{,}082$ upvotes, and posts on AI photo analysis and government tracking of insurance customers each exceeded $2{,}500$ -- elevating volume through February before it receded toward baseline (and, by March, below the preceding autumn's level) as attention moved on. Neither episode coincides with anything OSA-related. (r/privacy's role in the later Age Verification increase is covered separately below.)

\para{What the pre-existing Politics trends actually are.} Unlike the VPN pattern above, the Politics placebo-date failure at Royal Assent is not a data artefact or off-topic noise: reading the classified-relevant submissions directly, the rising volume in the two months before 26 October 2023 is substantively about the Online Safety Bill itself. r/ukpolitics submissions in this window include ``UK Online Safety Bill Will Mandate Dangerous Age Verification for Much of the Web'' (5 September), ``Britain Admits Defeat in Controversial Fight to Break Encryption'' and ``UK Government Drops `Spy Clause' from Online Safety Bill Amid Encryption Concerns'' (6 September), coverage of the Bill's final Parliamentary passage on 19 September (``Online Safety Bill: Crackdown on harmful social media content agreed'', ``Internet Censorship Galore as UK Online Safety Bill Becomes Law'', ``UK's Online Safety Bill finally passed by parliament''), and continuous coverage through to Royal Assent itself (``Online Safety Bill could become law on Thursday, Ofcom boss says'', 25 October; ``Beefed up internet rules become law'', 26 October; ``Tech firms cite risk to end-to-end encryption as Online Safety Bill gets royal assent'', 27 October). The Bill's final Commons vote on 19 September, more than a month before Royal Assent, generated sustained coverage that the placebo-date test correctly identifies as a pre-existing trend: for the Politics community, Royal Assent is the terminal event of a months-long, continuously-covered legislative process rather than a discrete surprise, which is exactly why it cannot be treated as a clean intervention for this series.

Enforcement shows the same pattern, and the placebo-date test's own diagnosis -- that the three-months-earlier date exceeds Enforcement's estimate -- points to a specific, real event. Ofcom published its Illegal Harms Codes of Practice on 16 December 2024, almost exactly three months before Enforcement (17 March 2025), generating an immediate cluster of r/ukpolitics and r/UnitedKingdom submissions (``New Ofcom Internet Regulations \& Online Safety Act'', ``Ofcom Publish New Online UK Internet Safety Codes of Practice'', ``Social media platforms have work to do to comply with Online Safety Act, says Ofcom'', all 16--17 December). Coverage continues without a clear break through January and February 2025 -- age-verification requirements for pornography sites (``All porn sites must `robustly' verify UK user ages by July'', 16 January), and small-site closures attributed directly to compliance burden (a cycling forum with 60,000 users announced it would shut down on 16 March 2025, the day before Enforcement, citing the Act by name) -- up to Enforcement itself (``Ofcom to enforce crackdown on illegal content across UK social media'', 16 March; ``What is the Online Safety Act? Law explained as it finally comes into force'', 17 March). As with Royal Assent, Enforcement is a landmark within a continuously-covered regulatory rollout rather than a discrete surprise, and the December Codes-of-Practice announcement is a concrete, dateable reason the placebo three months prior outperforms the milestone itself.

The same subreddit is also the largest single contributor to the Age Verification increase. Table~\ref{tab:av_subreddit_vpn} shows r/privacy alone accounts for $48.2\%$ of the total post-milestone rise in VPN-relevant content, more than all VPN-provider subreddits (VPN, ProtonVPN, MullvadVPN, nordvpn, Surfshark, Windscribe, PrivateInternetAccess) combined ($32.9\%$); reading the post-milestone submissions directly confirms this is substantive engagement with the milestone itself (e.g. ``Ahead of the UK ID verification laws I started using a VPN to bypass Reddit restrictions'', ``New UK Proof of Age requirements online -- Bypass with VPN''), not a classifier artefact. Table~\ref{tab:av_subreddit_politics} shows the Politics-side increase is even more concentrated, with r/ukpolitics and r/UnitedKingdom together accounting for $78.1\%$. Reading the classified-relevant submissions in these two subreddits directly confirms the same for Politics: 85.6\% explicitly mention OSA/Ofcom/age-verification terms (e.g. ``UK enforces age checks for adult content -- What you need to know'', ``A popular VPN is seeing a 1,400\% spike in signups as the UK's age verification law takes effect''), confirming this is substantive engagement rather than a classifier artefact.

\begin{table}[tb]
\centering
\caption{VPN-relevant content by subreddit, 8 weeks before vs.\ 8 weeks after Age Verification (25~Jul~2025), using the paper's high-precision classifier. Share is each subreddit's share of the total positive increase summed across subreddits.}
\label{tab:av_subreddit_vpn}
\resizebox{\columnwidth}{!}{%
\begin{tabular}{>{\raggedright\arraybackslash}p{0.3\columnwidth}rrrr}
\toprule
\textbf{Subreddit} & \textbf{Pre (8wk)} & \textbf{Post (8wk)} & \textbf{$\Delta$} & \textbf{Share} \\
\midrule
r/privacy & 406 & 1{,}456 & $+1{,}050$ & $48.2\%$ \\
r/degoogle & 73 & 456 & $+383$ & $17.6\%$ \\
r/VPN & 116 & 431 & $+315$ & $14.4\%$ \\
r/ProtonVPN & 130 & 304 & $+174$ & $8.0\%$ \\
r/MullvadVPN & 49 & 171 & $+122$ & $5.6\%$ \\
r/nordvpn & 26 & 118 & $+92$ & $4.2\%$ \\
r/Windscribe & 16 & 27 & $+11$ & $0.5\%$ \\
r/Surfshark & 16 & 24 & $+8$ & $0.4\%$ \\
r/censorship & 3 & 20 & $+17$ & $0.8\%$ \\
r/PrivateInternetAccess & 22 & 17 & $-5$ & $-0.2\%$ \\
r/PrivacyGuides & 0 & 8 & $+8$ & $0.4\%$ \\
\bottomrule
\end{tabular}}
\end{table}

\begin{table}[tb]
\centering
\caption{Politics-relevant content by subreddit, 8 weeks before vs.\ 8 weeks after Age Verification (25~Jul~2025). Share is each subreddit's share of the total positive increase summed across subreddits.}
\label{tab:av_subreddit_politics}
\resizebox{\columnwidth}{!}{%
\begin{tabular}{>{\raggedright\arraybackslash}p{0.3\columnwidth}rrrr}
\toprule
\textbf{Subreddit} & \textbf{Pre (8wk)} & \textbf{Post (8wk)} & \textbf{$\Delta$} & \textbf{Share} \\
\midrule
r/ukpolitics & 1{,}261 & 9{,}223 & $+7{,}962$ & $51.7\%$ \\
r/UnitedKingdom & 1{,}539 & 5{,}607 & $+4{,}068$ & $26.4\%$ \\
r/LabourUK & 126 & 1{,}174 & $+1{,}048$ & $6.8\%$ \\
r/AskUK & 890 & 1{,}893 & $+1{,}003$ & $6.5\%$ \\
r/Scotland & 153 & 498 & $+345$ & $2.2\%$ \\
r/LibDem & 27 & 228 & $+201$ & $1.3\%$ \\
r/NorthernIreland & 86 & 281 & $+195$ & $1.3\%$ \\
r/LegalAdviceUK & 109 & 233 & $+124$ & $0.8\%$ \\
r/GreenAndPleasant & 74 & 179 & $+105$ & $0.7\%$ \\
r/BritishProblems & 138 & 227 & $+89$ & $0.6\%$ \\
r/FreeSpeech & 28 & 111 & $+83$ & $0.5\%$ \\
r/CasualUK & 249 & 310 & $+61$ & $0.4\%$ \\
r/UKGreens & 3 & 55 & $+52$ & $0.3\%$ \\
r/labour & 0 & 35 & $+35$ & $0.2\%$ \\
r/Wales & 2 & 21 & $+19$ & $0.1\%$ \\
r/Tories & 0 & 9 & $+9$ & $0.1\%$ \\
r/England & 0 & 8 & $+8$ & $0.1\%$ \\
r/uklaw & 21 & 19 & $-2$ & $-0.0\%$ \\
\bottomrule
\end{tabular}}
\end{table}

\subsection{Age Verification as a Discrete Intervention}
\label{appendix:age_verification_manipulation_check}

Unlike Royal Assent, the Age Verification milestone corresponds to an
observable, dateable behavioural shock: major UK platforms began enforcing
age checks within a tight window around the regulatory deadline. Ofcom's
enforcement deadline was 25 July 2025~\cite{ofcom2025agechecksinforce}.
Independently reported platform compliance dates cluster tightly around it:
Discord from 21 July~\cite{discord2025osa}, Aylo (Pornhub/YouPorn/RedTube)
from 24 July~\cite{aylo2025ageassurance}, and Reddit's checks became binding
for UK users on 25 July~\cite{eff2025redditrollout}; Bluesky confirmed
changes were live by 25 July~\cite{forbes2025bluesky}. All four dates fall
within a single week of our weekly time series, so sub-week heterogeneity in
platform rollout timing is absorbed by our unit of analysis rather than
threatening identification.

To confirm the causal estimate is not an artefact of which exact day within
this cluster is used as the treatment date, we re-ran CausalImpact
(no-covariate spec) with the treatment date set to each of the four reported
compliance dates, for both VPN and Politics (Table~\ref{tab:app_placebo_ra}c,
bold rows). The estimated effect is stable across the entire 21--26 July
window within each group (VPN $+1248\%$ to $+1261\%$; Politics $+1470\%$ to
$+1490\%$; all $p{=}0.001$): all four dates fall within the
same weekly bin, so they map to an identical pre/post boundary and the small
residual variation reflects BSTS's stochastic fitting rather than a true date
effect. Treating 24/25 July as a single discrete intervention therefore does
not bias the estimate.

\para{Placebo test for Age Verification.}
We ran the same pre-trend placebo test used for Politics/Royal Assent, applied
instead to UK-authored VPN volume ahead of Age Verification (Table~\ref{tab:app_placebo_ra}c).
With fake intervention dates 8--12 weeks before 24 July 2025, and the post-window
bounded so it cannot overlap the real post-milestone rise, the placebo dates are
null at 10 and 12 weeks out and only marginally significant at 8 weeks out
($+41\%$, $p{=}0.083$), plausibly reflecting last-minute anticipatory activity.
This is over an order of magnitude smaller than the post-milestone effect
($+1248\%$ to $+1261\%$) and confined to the final two months at most, in
clear contrast to the systematic two-month pre-rise found for Politics ahead
of Royal Assent. We therefore treat Age Verification as a clean, discrete
intervention for VPN discourse, with at most a small anticipatory effect in
the two months before the deadline.

\para{Robustness to the Gaussian observation-noise assumption.}
\label{appendix:gaussian_robustness}
{The
UK-authored classified VPN series is a low-count series (pre-period mean
$\approx$7 documents/week, several weeks in the low single digits), for
which CausalImpact's default Gaussian observation model is an approximation
rather than an exact fit to discrete, right-skewed count data. To test
whether this affects our conclusion, we re-ran the identical specification
(growing pre-period, 8-week post-window, no covariate) on a
variance-stabilizing $\log(1+x)$ transform of the raw weekly counts -- the
standard correction for this concern -- and back-transformed the
counterfactual to natural units. The transformed model remains fully
significant (posterior tail-area $p{=}0.0$, $100\%$ posterior probability of
a causal effect) and yields an observed post-period total of 805 documents
against a back-transformed counterfactual of 53.7 (relative effect
$+1399\%$), in the same order of magnitude as our headline Gaussian estimate
($+1265\%$), if anything larger. The finding is therefore stable across
both model specifications.}

\begin{table*}[t]
\centering
\small
\caption{CausalImpact relative effects on user counts: unique, new, and
cumulative users per week, all authors vs.\ likely UK-resident. Same primary
spec as Table~\ref{tab:content_effects} (\S\ref{subsec:primary_spec}).
$^{**}p<0.05$, $^{*}p<0.10$, no marker/${\approx}$: not significant.
$^{\ddagger}$ }
\label{tab:users}
\resizebox{\textwidth}{!}{%
\begin{tabular}{llrrrrrrrrr}
\toprule
& & \multicolumn{3}{c}{\textbf{Oct 2023}} & \multicolumn{3}{c}{\textbf{Mar 2025}} & \multicolumn{3}{c}{\textbf{Jul 2025}} \\
\cmidrule(lr){3-5} \cmidrule(lr){6-8} \cmidrule(lr){9-11}
\textbf{Group} & \textbf{Filter} & \textbf{Uniq.} & \textbf{New} & \textbf{Cum.} & \textbf{Uniq.} & \textbf{New} & \textbf{Cum.} & \textbf{Uniq.} & \textbf{New} & \textbf{Cum.} \\
\midrule
\multirow{9}{*}{VPN}
 & Unmatched (non-UK)$^{\ddagger}$ & $-10\%^{**}$ & ${\approx}+13\%$ & $+4\%^{**}$ & $-23\%^{**}$ & $-31\%^{**}$ & $+3\%^{**}$ & $+42\%^{**}$ & $+55\%^{**}$ & $+4\%^{**}$ \\
 & Unmatched$^{\ddagger}$ & $-9\%^{**}$ & ${\approx}+13\%$ & $+4\%^{**}$ & $-24\%^{**}$ & $-32\%^{**}$ & $+3\%^{**}$ & $+46\%^{**}$ & $+62\%^{**}$ & $+4\%^{**}$ \\
 & Unmatched (UK)$^{\ddagger}$ & ${\approx}{-3\%}$ & ${\approx}{-16\%}$ & $+4\%^{**}$ & $-16\%^{**}$ & ${\approx}{-6\%}$ & $+3\%^{**}$ & $+102\%^{**}$ & $+206\%^{**}$ & $+9\%^{**}$ \\
 & Raw & $-7\%^{**}$ & ${\approx}{-10\%}$ & $+4\%^{**}$ & $-22\%^{**}$ & $-30\%^{**}$ & $+3\%^{**}$ & $+47\%^{**}$ & $+63\%^{**}$ & $+4\%^{**}$ \\
 & Raw (UK) & ${\approx}{-1\%}$ & ${\approx}{-14\%}$ & $+4\%^{**}$ & $-17\%^{**}$ & ${\approx}{-6\%}$ & $+3\%^{**}$ & $+114\%^{**}$ & $+230\%^{**}$ & $+9\%^{**}$ \\
 & Matched & ${\approx}{-3\%}$ & ${\approx}{-4\%}$ & $+4\%^{**}$ & $-13\%^{**}$ & ${\approx}{-8\%}$ & $+3\%^{**}$ & $+67\%^{**}$ & $+80\%^{**}$ & $+5\%^{**}$ \\
 & Matched (UK) & ${\approx}+6\%$ & ${\approx}{-3\%}$ & $+4\%^{**}$ & ${\approx}{-2\%}$ & ${\approx}+4\%$ & $+3\%^{**}$ & $+225\%^{**}$ & $+335\%^{**}$ & $+12\%^{**}$ \\
 & Classified & $-17\%^{**}$ & $-22\%^{**}$ & $+3\%^{*}$ & ${\approx}{-13\%}$ & $-15\%^{*}$ & $+2\%^{**}$ & $+290\%^{**}$ & $+303\%^{**}$ & $+12\%^{**}$ \\
 & Classified (UK) & ${\approx}{-13\%}$ & ${\approx}{-16\%}$ & $+4\%^{**}$ & ${\approx}{-14\%}$ & ${\approx}{-23\%}$ & $+2\%^{**}$ & $+1073\%^{**}$ & $+1230\%^{**}$ & $+38\%^{**}$ \\
\midrule
\multirow{9}{*}{Politics}
 & Unmatched (non-UK)$^{\ddagger}$ & $-29\%^{**}$ & $-41\%^{**}$ & ${\approx}+2\%$ & ${\approx}{-2\%}$ & ${\approx}{-1\%}$ & $+2\%^{**}$ & ${\approx}{-1\%}$ & ${\approx}+2\%$ & $+2\%^{*}$ \\
 & Unmatched$^{\ddagger}$ & $-13\%^{**}$ & $-42\%^{**}$ & ${\approx}+2\%$ & ${\approx}{-1\%}$ & ${\approx}+6\%$ & $+2\%^{**}$ & $-6\%^{*}$ & ${\approx}+17\%$ & $+2\%^{*}$ \\
 & Unmatched (UK)$^{\ddagger}$ & ${\approx}{-6\%}$ & ${\approx}{-37\%}$ & ${\approx}+2\%$ & ${\approx}+1\%$ & ${\approx}{-90\%}$ & $+2\%^{*}$ & $-7\%^{**}$ & ${\approx}+91\%$ & ${\approx}+1\%$ \\
 & Raw & $-13\%^{**}$ & $-43\%^{**}$ & ${\approx}+2\%$ & ${\approx}{-1\%}$ & ${\approx}{-2\%}$ & $+2\%^{**}$ & ${\approx}{-6\%}$ & ${\approx}+19\%$ & $+2\%^{*}$ \\
 & Raw (UK) & ${\approx}{-5\%}$ & $-48\%^{*}$ & ${\approx}+2\%$ & ${\approx}+1\%$ & ${\approx}{-76\%}$ & $+2\%^{*}$ & $-7\%^{**}$ & ${\approx}+124\%$ & ${\approx}+2\%$ \\
 & Matched & ${\approx}{-6\%}$ & $-23\%^{**}$ & $+3\%^{**}$ & $+10\%^{**}$ & ${\approx}+6\%$ & $+2\%^{**}$ & $+80\%^{**}$ & $+59\%^{**}$ & $+5\%^{**}$ \\
 & Matched (UK) & ${\approx}{-3\%}$ & $-23\%^{**}$ & $+3\%^{**}$ & $+9\%^{*}$ & ${\approx}+5\%$ & $+2\%^{**}$ & $+67\%^{**}$ & $+30\%^{**}$ & $+4\%^{**}$ \\
 & Classified & $+24\%^{*}$ & ${\approx}+10\%$ & $+5\%^{**}$ & $+51\%^{**}$ & $+27\%^{*}$ & $+3\%^{**}$ & $+1032\%^{**}$ & $+922\%^{**}$ & $+35\%^{**}$ \\
 & Classified (UK) & $+28\%^{**}$ & ${\approx}+13\%$ & $+5\%^{**}$ & $+51\%^{**}$ & ${\approx}+25\%$ & $+3\%^{**}$ & $+962\%^{**}$ & $+863\%^{**}$ & $+33\%^{**}$ \\
\bottomrule
\end{tabular}}
\end{table*}

\begin{table*}[t]
\centering
\small
\caption{CausalImpact relative effects on content volume: complete placebo
(Unmatched), keyword-matched, and Raw/Classified breakdown, posts/comments/total
(Raw/Classified rows reproduce Table~\ref{tab:content_effects} as posts/comments/total,
for a self-contained reference alongside the placebo and keyword-matched rows). Same
primary spec (\S\ref{subsec:primary_spec}). ``Unmatched'' = content
matching no VPN/OSA keyword; ``Matched'' = keyword-matched, any classifier
verdict; (UK)/(non-UK) restrict by author residency. \rev{Total columns
(bold headers) for the Unmatched/Matched rows were recomputed directly from
the underlying corpus for this table; Posts/Comments columns and the
Raw/Classified Total values reproduce already-published figures unchanged.}
$^{**}p<0.05$,
$^{*}p<0.10$; no marker/${\approx}$: not significant. $^{\ddagger}$ placebo
series, with the OSA mechanism removed (by topic, geography, or both): a large
effect here would signal general subreddit growth rather than a regulatory
response. The Unmatched rows are the clean comparison to the Raw/Classified
baselines, identical except that the treated
content is excluded.
{$^{\dagger}$ VPN Classified (UK) Posts rests on very few observations (258 posts total across the study window, only 6 in the 8-week window after Royal Assent) and is unstable across reruns; we recommend treating it as unreliable and reading the combined Total column instead.}}
\label{tab:content_effects_full}

{\setlength{\tabcolsep}{2.5pt}
\resizebox{\textwidth}{!}{%
\begin{tabular}{>{\raggedright\arraybackslash}p{0.09\textwidth}>{\raggedright\arraybackslash}p{0.15\textwidth}|*{3}{>{\centering\arraybackslash}p{0.075\textwidth}}|*{3}{>{\centering\arraybackslash}p{0.075\textwidth}}|*{3}{>{\centering\arraybackslash}p{0.075\textwidth}}}
\toprule
& & \multicolumn{3}{c}{\textbf{Oct 2023}} & \multicolumn{3}{c}{\textbf{Mar 2025}} & \multicolumn{3}{c}{\textbf{Jul 2025}} \\
\cmidrule(lr){3-5} \cmidrule(lr){6-8} \cmidrule(lr){9-11}
\textbf{Group} & \textbf{Filter} & Posts & Comments & \textbf{Total} & Posts & Comments & \textbf{Total} & Posts & Comments & \textbf{Total} \\
\midrule
\multirow{9}{*}{VPN}
 & Unmatched (non-UK)$^{\ddagger}$ & $-9\%^{*}$ & $-16\%^{**}$ & \textbf{$-13\%^{**}$} & $-10\%^{**}$ & $-27\%^{**}$ & \textbf{$-30\%^{**}$} & $+40\%^{**}$ & $+61\%^{**}$ & \textbf{$+64\%^{**}$} \\
 & Unmatched$^{\ddagger}$ & $-10\%^{*}$ & $-15\%^{**}$ & \textbf{$-13\%^{**}$} & $-9\%^{**}$ & $-27\%^{**}$ & \textbf{$-30\%^{**}$} & $+40\%^{**}$ & $+62\%^{**}$ & \textbf{$+63\%^{**}$} \\
 & Unmatched (UK)$^{\ddagger}$ & $-17\%^{**}$ & ${\approx}{-8\%}$ & \textbf{${\approx}{-7\%}$} & ${\approx}{-1\%}$ & $-14\%^{**}$ & \textbf{$-14\%^{**}$} & $+45\%^{**}$ & $+131\%^{**}$ & \textbf{$+126\%^{**}$} \\
 & Raw & ${\approx}+1\%$ & $-15\%^{**}$ & \textbf{$-21\%^{**}$} & $-16\%^{**}$ & $-26\%^{**}$ & \textbf{$-21\%^{**}$} & $+37\%^{**}$ & $+67\%^{**}$ & \textbf{$+75\%^{**}$} \\
 & Raw (UK) & ${\approx}+1\%$ & ${\approx}{-7\%}$ & \textbf{${\approx}{-6\%}$} & ${\approx}{-2\%}$ & $-15\%^{**}$ & \textbf{$-15\%^{**}$} & $+74\%^{**}$ & $+151\%^{**}$ & \textbf{$+145\%^{**}$} \\
 & Matched & $+20\%^{**}$ & $-16\%^{**}$ & \textbf{$-8\%^{**}$} & $-16\%^{**}$ & ${\approx}+6\%$ & \textbf{${\approx}{-3\%}$} & $+46\%^{**}$ & $+96\%^{**}$ & \textbf{$+82\%^{**}$} \\
 & Matched (UK) & $+34\%^{**}$ & ${\approx}{-1\%}$ & \textbf{${\approx}+6\%$} & ${\approx}{-2\%}$ & ${\approx}{-1\%}$ & \textbf{${\approx}{-6\%}$} & $+123\%^{**}$ & $+310\%^{**}$ & \textbf{$+258\%^{**}$} \\
 & Classified & ${\approx}+5\%$ & $-24\%^{**}$ & \textbf{$-20\%^{**}$} & $-30\%^{**}$ & ${\approx}{-9\%}$ & \textbf{${\approx}{-12\%}$} & $+220\%^{**}$ & $+349\%^{**}$ & \textbf{$+328\%^{**}$} \\
 & Classified (UK) & ${\approx}+181\%${$^{\dagger}$} & ${\approx}{-14\%}$ & \textbf{${\approx}{-16\%}$} & $-76\%^{*}${$^{\dagger}$} & ${\approx}{-8\%}$ & \textbf{${\approx}{-16\%}$} & $+786\%^{**}$ & $+1332\%^{**}$ & \textbf{$+1265\%^{**}$} \\
\midrule
\multirow{9}{*}{UK Politics}
 & Unmatched (non-UK)$^{\ddagger}$ & ${\approx}+7\%$ & $-26\%^{**}$ & \textbf{$-25\%^{**}$} & $+9\%^{*}$ & ${\approx}{-3\%}$ & \textbf{${\approx}{-3\%}$} & $+7\%^{*}$ & ${\approx}+1\%$ & \textbf{${\approx}+2\%$} \\
 & Unmatched$^{\ddagger}$ & ${\approx}+3\%$ & $-12\%^{**}$ & \textbf{$-13\%^{**}$} & ${\approx}+3\%$ & ${\approx}{-3\%}$ & \textbf{${\approx}{-2\%}$} & ${\approx}+3\%$ & ${\approx}{-4\%}$ & \textbf{${\approx}{-4\%}$} \\
 & Unmatched (UK)$^{\ddagger}$ & ${\approx}+1\%$ & $-9\%^{*}$ & \textbf{$-9\%^{**}$} & $-6\%^{*}$ & ${\approx}{-2\%}$ & \textbf{${\approx}{-2\%}$} & ${\approx}{-4\%}$ & $-6\%^{*}$ & \textbf{$-6\%^{*}$} \\
 & Raw & ${\approx}+4\%$ & $-12\%^{**}$ & \textbf{${\approx}{-4\%}$} & ${\approx}+2\%$ & ${\approx}{-3\%}$ & \textbf{{${\approx}{-2\%}$}} & $+5\%^{*}$ & ${\approx}{-2\%}$ & \textbf{${\approx}{-7\%}$} \\
 & Raw (UK) & ${\approx}+2\%$ & $-8\%^{*}$ & \textbf{{${\approx}{-8\%}$}} & $-5\%^{*}$ & ${\approx}{-2\%}$ & \textbf{{${\approx}{-2\%}$}} & ${\approx}{-2\%}$ & ${\approx}{-4\%}$ & \textbf{{${\approx}{-4\%}$}} \\
 & Matched & $+21\%^{**}$ & $-8\%^{*}$ & \textbf{$-9\%^{*}$} & $+10\%^{*}$ & $+11\%^{**}$ & \textbf{$+10\%^{**}$} & $+141\%^{**}$ & $+139\%^{**}$ & \textbf{$+121\%^{**}$} \\
 & Matched (UK) & $+16\%^{*}$ & ${\approx}{-5\%}$ & \textbf{${\approx}{-6\%}$} & $+14\%^{*}$ & $+11\%^{**}$ & \textbf{$+10\%^{*}$} & $+164\%^{**}$ & $+123\%^{**}$ & \textbf{$+106\%^{**}$} \\
 & Classified & ${\approx}+40\%$ & ${\approx}+24\%$ & \textbf{${\approx}+25\%$} & $+65\%^{**}$ & $+65\%^{**}$ & \textbf{$+65\%^{**}$} & $+1423\%^{**}$ & $+1466\%^{**}$ & \textbf{$+1474\%^{**}$} \\
 & Classified (UK) & ${\approx}+35\%$ & ${\approx}+27\%$ & \textbf{${\approx}+27\%$} & $+68\%^{**}$ & $+60\%^{**}$ & \textbf{$+60\%^{**}$} & $+1698\%^{**}$ & $+1490\%^{**}$ & \textbf{$+1481\%^{**}$} \\
\bottomrule
\end{tabular}}}
\end{table*}

\para{The Unmatched/Matched rows show the effect sharpening as the content filter tightens.}
The three rows in Table~\ref{tab:content_effects_full} not in Table~\ref{tab:content_effects} --
Unmatched, Unmatched (UK), and Matched -- sit between ``no filter at all'' and
the Classified rows already discussed in the main text, and they show the
Age Verification effect climbing at each step rather than appearing only
once a filter is applied. For VPN, the total effect rises from
$+63\%^{**}$ (Unmatched) to $+82\%^{**}$ (Matched) to $+328\%^{**}$
(Classified) to $+1265\%^{**}$ (Classified, UK) -- a 20-fold increase in the
estimated effect size from the loosest to the tightest filter. For Politics
the pattern is sharper still: Unmatched is a statistical null ($\approx{-4\%}$)
while Matched is already $+121\%^{**}$, climbing to $+1474\%^{**}$ once the
classifier confirms relevance. In both groups the placebo-style Unmatched
rows -- content excluded by definition from the treated series -- stay small
or null, while each successive filter step concentrates the estimate further.
This monotonic gradient is what we would expect if the milestone drove
genuine engagement with OSA-relevant content specifically, and not if the
Classified-row effects were an artefact of a classifier that happened to
correlate with unrelated subreddit-wide activity.

\subsection{Alternative-Specification Placebo and Date-Sensitivity Checks}
\label{appendix:earlier_placebo}

Figure~\ref{fig:rq1_gt} and Table~\ref{tab:trends} (both main text) report the primary CausalImpact estimate for UK VPN Google Trends search interest; the alternative specifications follow below.

Tables~\ref{tab:app_placebo_trends}--\ref{tab:app_datesens_trends} report
independent Google Trends placebo and date-sensitivity checks, not reported
elsewhere, under a longer-window specification that uses all remaining post
milestone data rather than the 8-week window used in the main text. The
Google Trends series were re-fetched for
``vpn'' (and ``weather'' as a placebo-outcome negative control). Under this longer-window specification the no-covariate Age Verification effect on UK
Google Trends search interest is $+96.6\%$; under the 8-week specification
used throughout the paper the same outcome is $+147\%$ (Table~\ref{tab:trends}).
Unlike the Reddit VPN outcome (Table~\ref{tab:app_placebo_ra}c), the Google Trends placebo-date test shows no
significant pre-trend at any lag (Table~\ref{tab:app_placebo_trends}), and the
placebo-outcome check against ``weather'' search interest shows only a small,
marginal effect ($-10.7\%$, $p=0.09$), plausibly attributable to ordinary
seasonality. The Google Trends outcome is thus a cleaner placebo case than the
Reddit content.

\begin{table}[tb]
\caption{Placebo-date pre-trend test, UK Google Trends search interest for ``vpn'', fake intervention dates before Age Verification. All non-significant, in contrast to the Reddit VPN outcome (Table~\ref{tab:app_placebo_ra}c).}
\label{tab:app_placebo_trends}
\begin{tabular}{lrr}
\toprule
\textbf{Fake date} & \textbf{Weeks before} & \textbf{Effect} \\
\midrule
2025-05-01 & 12 & $-3.5\%$ ($p=0.13$) \\
2025-05-15 & 10 & $-3.7\%$ ($p=0.16$) \\
2025-05-29 & 8  & $-3.0\%$ ($p=0.21$) \\
2025-06-12 & 6  & $-2.1\%$ ($p=0.30$) \\
2025-06-26 & 4  & $-3.1\%$ ($p=0.28$) \\
\bottomrule
\end{tabular}
\end{table}

\begin{table}[tb]
\caption{Date-sensitivity test, UK Google Trends ``vpn'' search interest: effect size across the platform-compliance date cluster. Sanity check against the true 24 Jul 2025 date reproduces $+96.6\%$, matching the $+95\%$ reported under the longer-window specification; the current 8-week specification gives $+147\%$ (Table~\ref{tab:trends}).}
\label{tab:app_datesens_trends}
\begin{tabular}{lr}
\toprule
\textbf{Treatment date} & \textbf{Effect} \\
\midrule
21 Jul 2025 & $+97.0\%^{**}$ \\
24 Jul 2025 & $+97.0\%^{**}$ \\
25 Jul 2025 & $+97.0\%^{**}$ \\
26 Jul 2025 & $+96.9\%^{**}$ \\
\bottomrule
\end{tabular}
\end{table}

\subsection{Post-Window-Length Robustness (Classified Content)}
\label{appendix:window_robustness_main}

This table reports the total (posts+comments) relative effect under 4-, 8-,
and 12-week post-windows. The headline UK-authored VPN values at Age
Verification are summarised in Section~\ref{subsec:rq1_method}.

\begin{table}[tb]
\caption{Robustness of the total (posts+comments) relative effect to post-window length, same spec as Table~\ref{tab:content_effects} otherwise. Royal Assent is non-significant (occasionally negative) at every short window across all four series; Age Verification is significant at every window tested, including daily resolution (Appendix, not tabulated).}
\label{tab:window_robustness}
\resizebox{\columnwidth}{!}{%
\begin{tabular}{lrrrr}
\toprule
\textbf{Series / Milestone} & \textbf{4wk} & \textbf{8wk} & \textbf{12wk} \\
\midrule
VPN (all), Royal Assent      & $-24\%^{**}$ & $-20\%^{**}$ & ${\approx}{-12\%}$ \\
VPN (all), Enforcement       & ${\approx}{-10\%}$ & ${\approx}{-12\%}$ & ${\approx}{-12\%}$ \\
VPN (all), Age Verification  & $+448\%^{**}$ & $+328\%^{**}$ & $+253\%^{**}$ \\
\midrule
VPN (UK), Royal Assent       & ${\approx}{-3\%}$ & ${\approx}{-16\%}$ & ${\approx}{-4\%}$ \\
VPN (UK), Enforcement        & ${\approx}+14\%$ & ${\approx}{-16\%}$ & ${\approx}{-4\%}$ \\
VPN (UK), Age Verification   & $+1132\%^{**}$ & $+1265\%^{**}$ & $+932\%^{**}$ \\
\midrule
Politics (all), Royal Assent & ${\approx}+35\%$ & ${\approx}+25\%$ & ${\approx}+6\%$ \\
Politics (all), Enforcement  & $+56\%^{*}$ & $+65\%^{**}$ & $+40\%^{**}$ \\
Politics (all), Age Verification & $+2884\%^{**}$ & $+1474\%^{**}$ & $+1105\%^{**}$ \\
\midrule
Politics (UK), Royal Assent  & ${\approx}+31\%$ & ${\approx}+27\%$ & ${\approx}+9\%$ \\
Politics (UK), Enforcement   & $+57\%^{*}$ & $+60\%^{**}$ & $+39\%^{**}$ \\
Politics (UK), Age Verification & $+3534\%^{**}$ & $+1481\%^{**}$ & $+1115\%^{**}$ \\
\bottomrule
\end{tabular}}
\end{table}

\begin{table}[tb]
\caption{Effect size vs.\\ UK-residency confidence threshold $k$ (minimum number of distinct UK geo-subreddits an author must have posted in to be flagged UK-resident). UK-authored content, total posts+comments, 8-week window, no covariate. {``VPN Cls.''/``Politics Cls.'' rows use classified (OSA-keyword-relevant) content; ``VPN Raw''/``Politics Raw'' rows use all content in the subreddit group, unfiltered by the classifier.} $^{**}p<0.05$, $^{*}p<0.10$. \rev{We report $k{\leq}5$ only; beyond that the estimator becomes unstable due to low sample size (\S\ref{appendix:uk_author_robustness}).} {$n$ is the number of authors surviving each threshold in that row's estimation sample; it falls steeply, which is why we do not report $k{\geq}6$. M. Act.\ reports each threshold's mean per-author activity for that row's content definition (classified activity for VPN Cls./Politics Cls.; raw activity for VPN Raw/Politics Raw).}}
\label{tab:app_k_robustness}
\resizebox{\columnwidth}{!}{%
\begin{tabular}{lrrrrrr}
\toprule
\textbf{Group} & \textbf{$k$} & {\textbf{$n$}} & \textbf{Oct 2023} & \textbf{Mar 2025} & \textbf{Jul 2025} & {\textbf{M. Act.}} \\
\midrule
{VPN Raw} & {$\geq$1} & {24{,}482} & {$-6.5\%$} & {$-15.2\%^{**}$} & {$+144.3\%^{**}$} & {7.24} \\
{VPN Raw} & {$\geq$2} & {10{,}258} & {$+4.2\%$} & {$-11.8\%$} & {$+228.1\%^{**}$} & {6.86} \\
{VPN Raw} & {$\geq$3} & {6{,}082} & {$+5.6\%$} & {$-11.0\%$} & {$+270.9\%^{**}$} & {6.92} \\
{VPN Raw} & {$\geq$4} & {3{,}921} & {$+28.1\%^{**}$} & {$-19.5\%^{**}$} & {$+277.0\%^{**}$} & {7.18} \\
{VPN Raw} & {$\geq$5} & {2{,}659} & {$+43.1\%^{**}$} & {$-16.2\%^{**}$} & {$+328.8\%^{**}$} & {7.43} \\
{VPN Cls.} & $\geq$1 & {1{,}468} & $-15.6\%$ & $-14.8\%$ & $+1256.3\%^{**}$ & {0.085} \\
{VPN Cls.} & $\geq$2 & {776} & $+7.4\%$ & $-25.6\%$ & $+2102.4\%^{**}$ & {0.099} \\
{VPN Cls.} & $\geq$3 & {535} & $+32.6\%$ & $-4.0\%$ & $+2495.1\%^{**}$ & {0.113} \\
{VPN Cls.} & $\geq$4 & {384} & $+14.2\%$ & $+11.3\%$ & $+3149.4\%^{**}$ & {0.126} \\
{VPN Cls.} & $\geq$5 & {269} & $+56.5\%$ & $+3.7\%$ & $+3187.5\%^{**}$ & {0.134} \\
{Politics Raw} & {$\geq$1} & {646{,}370} & {$-8.2\%^{*}$} & {$-2.1\%$} & {$-4.1\%$} & {69.42} \\
{Politics Raw} & {$\geq$2} & {266{,}027} & {$-4.6\%$} & {$-0.2\%$} & {$-4.5\%$} & {126.04} \\
{Politics Raw} & {$\geq$3} & {128{,}875} & {$-1.7\%$} & {$+2.0\%$} & {$-5.8\%^{*}$} & {194.61} \\
{Politics Raw} & {$\geq$4} & {69{,}186} & {$+0.9\%$} & {$-3.2\%$} & {$-11.7\%^{**}$} & {272.95} \\
{Politics Raw} & {$\geq$5} & {39{,}362} & {$-6.4\%^{*}$} & {$-2.7\%$} & {$-12.0\%^{**}$} & {364.77} \\
{Politics Cls.} & $\geq$1 & {17{,}071} & $+26.0\%$ & $+58.8\%^{**}$ & $+1450.3\%^{**}$ & {0.062} \\
{Politics Cls.} & $\geq$2 & {12{,}523} & $+28.3\%$ & $+61.4\%^{**}$ & $+1482.8\%^{**}$ & {0.112} \\
{Politics Cls.} & $\geq$3 & {9{,}099} & $+32.3\%^{*}$ & $+59.0\%^{**}$ & $+1488.3\%^{**}$ & {0.171} \\
{Politics Cls.} & $\geq$4 & {6{,}622} & $+36.6\%^{*}$ & $+64.7\%^{**}$ & $+1340.9\%^{**}$ & {0.218} \\
{Politics Cls.} & $\geq$5 & {4{,}826} & $+47.2\%^{**}$ & $+58.5\%^{**}$ & $+1333.2\%^{**}$ & {0.279} \\
\bottomrule
\end{tabular}}
\end{table}

\subsection{UK-Resident Author Identification: Robustness Checks}
\label{appendix:uk_author_robustness}

\para{UK-resident author identification.} The UK-resident author filter is
built from a comprehensive per-document index (65{,}762{,}361 rows spanning
all 29 study subreddits, Jan~2021--Dec~2025, with no duplicate IDs and
complete field coverage), yielding 652{,}200 UK-resident authors. The UK-geo
subreddit set comprises the 37 subreddits listed in Table~\ref{tab:subreddits}.

\para{UK-affinity threshold sensitivity.} {Table~\ref{tab:app_k_robustness} reports} the Age Verification effect at
increasing minimum-evidence thresholds $k$ (the number of distinct UK
geo-subreddits an author must have posted in), up to $k{\leq}5$. Most flagged
UK authors (1{,}089{,}836 of 1{,}409{,}171 geo-active authors) meet only the
$k{\geq}1$ threshold. For VPN, the effect size increases monotonically with
$k$ over this range, from $+1256\%$ at $k{\geq}1$ to $+3188\%$ at $k{\geq}5$,
significant throughout -- the effect concentrates among the most
confidently-identified UK residents. We do not report thresholds beyond
$k{\geq}5$: only 269 VPN authors survive that threshold, and at $k{\geq}6$ and
above the estimator becomes unstable due to the low sample size and no longer
produces a reliable effect. Politics, whose smallest threshold here still
retains 4{,}826 authors, is stable across the whole reported range ($+1450\%$
at $k{\geq}1$ to $+1333\%$ at $k{\geq}5$), which is what the VPN series would
look like were sample size not the binding constraint.

\para{Ruling out a prolific-author confound.} {We employ $k$ as our confidence threshold for the Uk-resident label which denotes the
number of distinct UK geo-subreddits an author has posted in. Higher $k$ means
stronger evidence the author actually lives in the UK. A natural concern is
that raising $k$ instead (or additionally) selects for generally prolific
Reddit accounts, which could inflate the effect-size pattern above for
reasons unrelated to residency confidence. We test this by computing each author's raw
activity (all posts and comments in the relevant subreddit group) and
OSA-keyword-classified activity as $k$ increases (Table~\ref{tab:app_k_robustness}).
The two groups diverge. For VPN, where the
effect size climbs steadily with $k$ ($+1256\%$ to $+3188\%$), author
prolificacy is essentially flat (7.24 to 7.43 mean raw documents, Spearman
$\rho{=}0.01$; classified activity $\rho{=}0.04$): the rising effect cannot
be explained by recruiting heavier posters, and is better attributed to
increasing residency confidence itself. For Politics, author prolificacy
instead rises sharply over the same range (69.4 to 364.8 raw documents,
$\rho{=}0.45$; classified activity 0.062 to 0.279, $\rho{=}0.12$), yet the
effect size stays essentially flat and non-monotonic ($+1450\%$ to $+1483\%$
to $+1333\%$). If the Politics estimate were being driven by this
compositional shift toward more prolific posters, we would expect its effect
to rise with $k$ the way VPN's does; instead it holds steady despite the
shift, indicating the Politics result is not an artefact of the
prolific-author confound either, even though the confound is clearly
present in the underlying population there.}

\para{Posting-time validation.} As an independent, subreddit-participation-free
check on the geo-filter, UK-flagged authors show a lower share of activity
during UK nighttime hours (1--5am UTC) than non-UK-flagged authors in both
corpora (VPN: 9.3\% vs.\ 13.1\%; Politics: 3.4\% vs.\ 5.6\%), in the
theoretically expected direction. The gap is real but modest and should be read
as corroborating evidence.

\para{Placebo checks on the new-user rise.} Placebo series distinguish
genuine Age-Verification engagement from ordinary subreddit growth. For
Politics the placebo is clean: non-UK users of keyword-unmatched content show
no change at Age Verification (${\approx}{-1\%}$, $p{=}0.42$) while
relevant-content users rise roughly tenfold. For VPN it is less clean --
unmatched non-UK users also rise ($+42\%^{**}$) -- but this is about seven
times smaller than the all-author effect and twenty-five times smaller than
the UK-author one, limiting the interpretation without overturning it. A final
check fixes relevance and removes only the UK label: for VPN the non-UK series
sits well below the UK-author effect ($+216\%$ vs.\ $+1075\%$), as a
UK-specific mechanism predicts, whereas for Politics it slightly exceeds it
($+1190\%$ vs.\ $+971\%$) -- the clearest sign the Politics surge is shared
across UK and non-UK authors. That higher non-UK figure is explained next.

\para{Geo-proxy recall and the non-UK Politics figure.} Of the 9{,}012 authors
posting classified-relevant Politics content in the window, 6{,}608 ($73\%$)
were new to the series, and these newcomers are UK-flagged at $69.7\%$ against
$83.0\%$ for incumbents. Since UK-flagged authors are $75\%$ of the Politics
pool, that modest tilt inflates the smaller non-UK bucket disproportionately in
relative terms: the observed split predicts non-UK growing $1.33\times$ faster,
close to the $1.24\times$ observed. We therefore read the Politics non-UK figure
as an artefact of geo-proxy recall rather than a substantive finding.

\section{LLM Prompts}
\label{appendix:prompts}

\subsection{Relevance Classification -- VPN Subreddits}
\begin{flushleft}\small\itshape
You are classifying Reddit posts from VPN and privacy subreddits for a study on how the UK Online Safety Act affects VPN usage.

Mark as relevant (1) ONLY if the post explicitly discusses: government regulation, censorship, or surveillance as a reason to use a VPN; age verification, content blocks, or online safety laws (any country); privacy rights, civil liberties, or anonymity in a political or regulatory context; or ISP blocking, DNS filtering, or state-level internet restrictions.

Mark as irrelevant (0) if the post is about: which VPN is fastest/cheapest/best for streaming or gaming; torrenting or bypassing geo-restrictions for entertainment; technical setup help with no privacy/regulation context; VPN pricing, deals, or affiliate recommendations; or general security questions not tied to regulation or surveillance.

If the post does not clearly fit a relevant category, mark as irrelevant (0). Respond with ONLY a JSON array of 1s and 0s, one per post, in the same order. Example for 3 posts: [1, 0, 1]. No explanation, no other text.
\end{flushleft}

\subsection{Relevance Classification -- UK Politics Subreddits}
\begin{flushleft}\small\itshape
You are classifying Reddit posts from UK politics subreddits for a study on how the UK Online Safety Act affects VPN and privacy discourse.

Mark as relevant (1) ONLY if the post explicitly discusses: the UK Online Safety Act, Online Safety Bill, or Ofcom regulation; VPNs, proxies, or circumvention tools in a UK context; age verification, content blocking, or internet censorship in the UK; or encryption policy, surveillance, or privacy rights in a UK regulatory context.

Mark as irrelevant (0) if the post is about: NHS, housing, cost of living, immigration, or other non-internet policy; elections, political parties, or politicians unrelated to internet regulation; non-UK politics or international affairs; general tech news with no UK regulatory angle; personal legal advice about online conduct; or illegal online content questions unrelated to OSA enforcement.

If the post does not clearly fit a relevant category, mark as irrelevant (0). Respond with ONLY a JSON array of 1s and 0s, one per post, in the same order. Example for 3 posts: [1, 0, 1]. No explanation, no other text.
\end{flushleft}

\subsection{Sentiment Classification}
\begin{flushleft}\small\itshape
You are classifying Reddit posts and comments about the UK Online Safety Act (OSA) for a research study. For each post, assign a sentiment label based on the author's attitude toward the OSA and related privacy/regulation topics: \textbf{negative} (critical, opposed, frustrated, or concerned about the OSA, age verification, government surveillance, or internet censorship); \textbf{neutral} (informational, factual, or no clear opinion); or \textbf{positive} (supports the OSA, age verification, or related regulatory measures).

Respond with ONLY a JSON array of labels, one per post, in the same order. Example for 3 posts: [``negative'', ``neutral'', ``positive'']. No explanation, no other text.
\end{flushleft}

\subsection{Document Summarisation}
\begin{flushleft}
\small\itshape You are analysing Reddit posts from UK users discussing the UK Online Safety Act (OSA).
Below are posts from a single discussion cluster. Summarise
1. What specific concerns or arguments do users raise?
2. What examples, analogies, or real-world cases do they cite?
3. What is the overall tone and framing?
Be concrete and specific. Quote or paraphrase specific points where useful.
\end{flushleft}

\subsection{Keyword Lists}
\label{appendix:keywords}

Documents were flagged as potentially OSA- or VPN-related using the
case-insensitive substring keyword lists below. Keyword
matching served only as a permissive first-pass filter to bound the
candidate set; final relevance was determined by the classifier, not by
keyword presence.

\paragraph{OSA keywords (27 terms).}
\texttt{encryption}, \texttt{age}, \texttt{porn}, \texttt{pron},
\texttt{facial recognition}, \texttt{age verification},
\texttt{id verification}, \texttt{online safety act},
\texttt{identity verification}, \texttt{illegal content},
\texttt{face scan}, \texttt{osa}, \texttt{online safety bill},
\texttt{ofcom}, \texttt{id check}, \texttt{harmful content},
\texttt{age check}, \texttt{site blocking}, \texttt{yoti},
\texttt{face verification}, \texttt{age assurance}, \texttt{geo-block},
\texttt{internet law}, \texttt{veriff}, \texttt{onfido},
\texttt{prove age}, \texttt{dcms}.

\paragraph{VPN keywords (44 terms).}
\texttt{vpn}, \texttt{proxy}, \texttt{bypass}, \texttt{anonymity},
\texttt{tor}, \texttt{nord}, \texttt{workaround}, \texttt{dns},
\texttt{proton}, \texttt{nordvpn}, \texttt{mullvad}, \texttt{cloudflare},
\texttt{surfshark}, \texttt{mozilla}, \texttt{adguard},
\texttt{expressvpn}, \texttt{windscribe}, \texttt{private internet access},
\texttt{anonymous browsing}, \texttt{virtual private network},
\texttt{cyberghost}, \texttt{ivpn}, \texttt{airvpn}, \texttt{tunnelbear},
\texttt{purevpn}, \texttt{keepsafe}, \texttt{astrill}, \texttt{smart dns},
\texttt{privado}, \texttt{ipvanish}, \texttt{socks5}, \texttt{psiphon},
\texttt{torguard}, \texttt{hide.me}, \texttt{speedify}, \texttt{browsec},
\texttt{ovpn}, \texttt{total vpn}, \texttt{shellfire}, \texttt{privatevpn},
\texttt{vyprvpn}, \texttt{ivacy}, \texttt{x-vpn}, \texttt{privadovpn}.

\end{document}